\documentclass[twocolumn,pra,superscriptaddress]{revtex4-2}

\usepackage{amsmath,mathtools}
\usepackage{subfigure}
\usepackage{hyperref}

\usepackage{physics}
\newcommand{\mi}{\mathrm{i}}
\newcommand{\fleq}[1]{Eq.~\eqref{#1}}
\usepackage{diagbox}

\begin{document}

\title{Quantum non-Gaussian high Fock states of light pulses and their superpositions}

\author{G.~P.~Teja}
\email[]{teja4477@gmail.com}
\affiliation{Department of Optics, Faculty of Science, Palacký University, 771 46 Olomouc, Czech Republic}
\author{Chandan Kumar}
\affiliation{The Institute of Mathematical Sciences, C.I.T. Campus, Taramani, Chennai, India 600113}
\affiliation{Homi Bhabha National Institute, Training School Complex, Anushakti Nagar, Mumbai 400085, India.}
\author{Luk\'{a}\v{s} Lachman}
\affiliation{Department of Optics, Faculty of Science, Palacký University, 771 46 Olomouc, Czech Republic}
\affiliation{Laboratoire Kastler Brossel, Sorbonne Université, CNRS, Collége de France, 75005 Paris, France}
\author{Radim Filip}
\email[]{filip@optics.upol.cz}
\affiliation{Department of Optics, Faculty of Science, Palacký University, 771 46 Olomouc, Czech Republic}
\begin{abstract}
The generation of high Fock states of light pulses and their superpositions with provable quantum non-Gaussian features is still very challenging, although the power of conditional methods to herald the approximate state from the available Gaussian states is growing. The atom-light interaction in the high-Q cavity has been considered a viable alternative to the heralded Fock states of the light pulses from nonlinear optics limited to three-photon Fock states for the last decade.
Here, by optimizing the realistic protocol combining it with available optical delay elements we conclusively predict filtering of Fock states up to ten photons with a high success rate of $20\%$ using a hierarchy of quantum non-Gaussian criteria. 
Moreover, the filtering protocol enables the preparation of superposition of Fock states and we analyse this emerging case up to two photons with with provable quantum non-Gaussian coherence and high success rate $50\%$.
To demonstrate their quality for applications, we evaluate the robustness of such features, the bunching capability in a linear network, and the sensing capability to estimate the magnitude of unknown force, noise and phase, 
These assessments outline the essential conditions and application criteria for realistic optical cavity QED interaction on light pulses to outperform photon detection methods.
\end{abstract}

\maketitle
\pretolerance=1000
\section{Introduction}
Quantum optics and photonic technology developed broadly over the last decade, mainly increasing the number of entangled and interfering modes in Gaussian states of light. It allowed the experimental test of Gaussian boson sampling \cite{boson2019,Zhong_2020,phase2021}, the construction of Gaussian cluster states for measurement-induced quantum computing \cite{Yokoyama_2013,Yoshikawa_2016,2019_warit,Larsen_2019}, the broad use of Gaussian squeezed light for quantum sensing \cite{2010_wolf,Guo_2019,yap2020,casacio2021,Xia_2023} and quantum key distribution \cite{madsen2012,gehring2015,Walk_2016,Kovalenko-21}. However, to investigate unexplored physics beyond Gaussian states and their classical mixtures and advance the applications further and turn them into more universal \cite{lloyd1999} and fault-tolerant forms \cite{2001gott,2016_mich}, either quantum non-linearities beyond the quadratic or quantum non-Gaussian states need to be involved.

In quantum optics with Gaussian states, nonlinear single-photon detection typically provides such quantum non-Gaussianity. A low-noise recognition of any number of non-zero photons from a vacuum allows a trigger for conditional single photon states from a split single-mode squeezed light \cite{2006_alexei,2006_nee,2009_our}, two-mode squeezed light \cite{lvov_2001,Takaha_2008,Bimbard_2010,Takahashi_2010,Yoshik_2013} or interfering multimode squeezed states of light \cite{Lee_2011,Takeda_2013,Morin_2014,Ra_2019,Giovanni_2020,Darras_2022}. Such conditional states limitedly exhibit quantum non-Gaussian aspects. It happens even if the trigger efficiency is low and collection and verification efficiency is limited, however, with an extremely low probability of success.
When the probability of such a heralded single photon generation overcomes 50$\%$, the single photon state further exhibits a negative Wigner function \cite{schleich2011}. This can be measured directly using unbalanced homodyning \cite{1999_Ban,Laiho_2009} or eventually estimated indirectly from homodyne tomography \cite{2009_Lvov}.

Negativity of Wigner function is a necessary condition for universal quantum processing \cite{2012_Mari,Veitch_2012}. 
However, the negativity is insufficient for already simpler effects and applications, and for breeding of more complex states \cite{Shunya_2024}. Therefore, further hierarchical criteria of quantum non-Gaussianity and subsequent analysis are highly required \cite{Lachman_2019,2021_Zapl}.

Recent approach is to use low-noise multiplexed single-photon detectors \cite{Cooper_2013,Yukawa_2013} or, ideally, photon-number resolving detectors (PNRDs) \cite{Endo_2023,sonoyama2024}. However, sensitivity to inefficiency of the detectors in the detection, and noise, as well as thresholds for hierarchical criteria increases \cite{Lachman_2019}. 
Simultaneously, the heralding rate does not increase sufficiently high.
Therefore, for a decade, Fock state generation is practically limited to the preparation of three-photon states witnesses by three annuli of the negative Wigner function \cite{Cooper_2013,Yukawa_2013}. 
Moreover, for the superposition up to two-photon Fock states of light, more fragile quantum coherences represented by the individual off-diagonal elements in the Fock basis are still not ultimately quantum non-Gaussian.

This bottleneck needs to be overcome to fully explore the photonic world and applications mentioned above, namely in quantum sensing \cite{Oh_2020,Han_2023}, quantum error correction \cite{Shunya_2024}, non-Clifford gates for quantum computing \cite{2021_Konno} and non-linear optical measurements \cite{Sakaguchi_2023}. Recently, new attempts to overcome this limiting bottleneck have been presented using improved generation and detection techniques \cite{takase2024,2024_Winnel,2023_Takase}. However, such generations still use only linearized couplings between photons, and all nonlinearity is provided only in the triggering detection.

A viable approach can be based on optical interaction with atomic, molecular and solid-state emitter levels \cite{2020_meyer,Slussarenko_2019,reiser_2015}. Single-photon light deterministically generated from a single emitter in a free space can already be quantum non-Gaussian \cite{Higginbottom_2016,liu_2024,checchinato2024}. However, it does not exhibit a negative Wigner function due to optical losses. It is also hard to redirect such emitted light in free space to the same or another emitter and use stimulated emission for higher photon generation in a well-defined mode. It complicates even more if many free space emitters should contribute coherently to such protocols. Therefore, a prospective solution is to use a cascaded interaction with a single emitter \cite{Daiss_2021,Langenfeld_2021,Brekenfeld_2020} or even more emitters \cite{2014_Badshah,Thomas_2024,Hartung_2024} in the high-quality optical cavity. 
For optical frequencies, such interaction is either a strong and resonant Jaynes–Cummings (JC) type within rotating wave approximation \cite{reiser_2015} or a weak, largely off-resonant of the dispersive type \cite{Reiserer_2013}. The dispersive protocols have been experimentally tested \cite{2019_Daiss,Hacker_2019} and further elaborated \cite{Verma_2022}. 
Conversely, superconducting systems with an intra-cavity microwave states can employ different ECD and SNAP gates 
\cite{Eickbusch_2022, Kudra_2022} or pulse engineering \cite{Heeres_2017, krisnand_2025, krisnanda2025}
beyond dispersive interaction to achieve fully deterministic state preparation.

Here, we utilize the emitter in the optical cavity not just for heralding but dominantly for high-fidelity filtering of the desired photon numbers of optical pulses and surprisingly, their binary superpositions. We keep such conditional preparation of the Fock states and superpositions from the initial Gaussian states of the pulses for its advantages. However, we use the power of optical emitters and their dynamics in a cavity
%On the other hand, in optical experiments, generating higher Fock states with a conclusive and high quantum non-Gaussian rank remains a challenging task. 
%However, the idealized protocol suggested in \cite{teja2023} cannot predict the quality of such generated Fock states without necessary criteria and is not extended to their binary superpositions.
%Towards experimental tests, here
to extend the cavity QED experiment \cite{Hacker_2019} by using the modern tools of optical delays \cite{yutaro_2021,Yonezu-2023} to significantly outperform photon detection methods on Gaussian states.

Recent work proposed a repetitive single-atom protocol for distilling high-Fock states to explore the limits of hybrid optical technologies~\cite{teja2023}, with a superconducting experiment demonstrating the feasibility of an intracavity Fock-state filtration \cite{deng2023}.
Conversely, in optical pulse experiments, generating higher Fock states and binary Fock-state superpositions remains a significant challenge.
Moreover, an essential evaluation of the quantum non-Gaussian (QNG) rank—a definitive hierarchical measure—is needed to rigorously justify claims of enhanced quantum non-Gaussianity. Crucially, while Wigner negativities observed in the experiment \cite{deng2023} serve as a witness of nonclassicality, they do not certify the QNG rank, as such negativities persist in principle for any loss below $50\%$. On the other hand, the QNG rank is loss-sensitive \cite{Lachman_2019}.
Furthermore, Wigner function negativities provide no direct insight into quantum coherences governed by off-diagonal elements in the Fock basis. 
%Additionally, superconducting systems can utilize a variety of more powerful interactions and control mechanisms beyond dispersive coupling, enabling more deterministic state preparation.

For the first time, we employ definitive hierarchical criteria to rigorously evaluate the quantum non-Gaussian (QNG) rank of Fock states and their superpositions generated in optical cavity QED, encompassing both diagonal and off-diagonal elements in the Fock basis. This approach enables us to predict the achievable limits of QNG features for states up to ten photons while also paving the way for binary superpositions. 
To fully understand the prospects of optical cavity QED for that, we compare a fully conclusive hierarchy of measures, including the QNG rank \cite{Lachman_2019}, quantum non-Gaussian depth \cite{Straka_2014}, photon bunching signatures \cite{2021_Zapl}, and force sensing applicability \cite{podhora2022}.

Based on these criteria and measures, we carefully optimize the most feasible sequential configuration of the pulsed optical cavity QED experiment \cite{Hacker_2019} to a realistic protocol, implementable in the the current laboratories. 
Our evaluations establishes the experimental conditions required to produce ten-photon Fock states of the optical pulses with a corresponding QNG rank and a probability of 20\%. Therefore, it surpasses Gaussian methods using photon-number resolving detectors. 
Moreover as an immediate extension, a binary superpositions of up to two photons with a genuine QNG coherence and probability 50\% is proposed to enhance the Gaussian method with photon detectors \cite{asenbeck2025}. 
Therefore, their experimental verification will provide full access to the quantum non-Gaussian regime in quantum optics, on par with state-of-the-art mechanical and microwave experiments \cite{rahman2025,krisnanda2025,kov2024} that already verified high QNG rank of Fock states and their superposition, conclusively. 
All these quantitative results \cite{podhora2022}, confirm the feasibility of high Fock state generation of optical pulses in a delay-based optical cavity-QED and open further investigation towards their superpositions and applications in quantum sensing, as well as future uses in quantum error correction and quantum computing.

\section{Evaluation AND APPLICATION}
\subsection*{QNG-hierarchy and QNG-depth}
A pure state is said to show genuine $n$-photon quantum non-Gaussianity (QNG) if it cannot be expressed as a Gaussian transformation of a
superposition of states containing up to $n-1$ photons, i.e. \cite{lachman2019}
\begin{align}\label{eq:qng}
\Psi_{n} \neq \hat{\mathcal{D}}(\alpha)  \hat{\mathcal{S}}(r)  \psi_{n-1}
~ \text{with} ~ \psi_{n-1} = \sum_{k}^{n-1} c_k \ket{k},
\end{align}
where $\hat{\mathcal{S}}(r) = \exp[\frac{1}{2}(r^* \hat{a}^2 - r \hat{a}^{\dagger 2})]$  and $\hat{\mathcal{D}}(\alpha)=\exp(\alpha \hat{a}^\dagger- \alpha^* \hat{a})$ 
are the  general single-mode squeezing and displacement operators, respectively.  Genuine $n$-photon QNG of impure states also rejects any processes that affect the amplitude $\alpha$, squeezing $r$ and amplitudes $c_k$ of the core state $\psi_{n-1}$ in Eq.~(\ref{eq:qng}). 
We certify the genuine $n$-photon QNG for a mixed state $\rho$ using a sufficient condition: the probability $p_n(\rho) = \langle n|\rho|n\rangle$ must satisfy $p_n > \mathcal{T}_n$, where the threshold
\begin{equation} \label{eq:rqng}
	\mathcal{T}_n = \max_{\alpha,r,c_k}  \abs{ \langle n| {\hat{\mathcal{D}}(\alpha)  \hat{\mathcal{S}}(r) \psi_{n-1}}}^2 ,
\end{equation}
covers values of the probability $p_n$ that any Gaussian evolution of the state $\Psi_n$ induces. For example numerical calculations show that   $\mathcal{T}_{10} = 0.65$ \cite{podhora2022}.
For $n>1$, thresholds are above the conditions to observe the negativity of the Wigner function.

For cases, when we fail in achieving $p_n >\mathcal{T}_n$, we use a relative but still hierarchical criterion (r-QNG) considering the probability of the multi-photon errors $p_{{\scriptstyle >n}}=\sum_{k=n+1}^{\infty} \langle k|\rho|k\rangle$ \cite{Lachman_2019}.  Fig.~\ref{fig:dep} depicts example of such a threshold derived for the probabilities $p_{10}$ and $p_{ {\scriptstyle >10}}$. It shows that the values of $p_{10}$ drops substantially for small values of $p_{{\scriptstyle >10}}$, and therefore the criterion certifies genuine $n$-photon r-QNG of the filtered states.
%with strongly suppressed error probability $p_{ {\scriptstyle >10}}\ll 1$.

Detecting genuine quantum non-Gaussian features using this relative criterion under photon loss is advantageous, which is the main imperfection of optical experiments.
To explore the correlation between QNG, r-QNG and noise, we investigate the depth of tolerable loss, following the approach  used for non-classical optical states \cite{Straka_2014}.
The additional attenuation testing the robustness of QNG and r-QNG features is modeled as:
\begin{align}\label{eq:dep}
p'_n =  \sum_{m=n}^\infty \binom{m}{n} T^n (1-T)^{m-n} p_m,
\end{align}
here, $p_m=\ev{\rho}{m}$ and $T$ is the transmittance of the noisy channel. 
The QNG depth (r-QNG depth) is defined as amount of $T$ tolerable until $p_{n}$ falls below the QNG-threshold $\mathcal{T}_n$. This methodology allows us to compare quantum non-Gaussian states conclusively by their robustness. 

To access the quantum non-Gaussian coherences, a criterion is developed for binary superpositions of the Fock states of the form $\frac{1}{\sqrt{2}}(\ket{n_1} + e^{i\phi}\ket{n_2})$ \cite{kov2024,lachman2025}. The coherence measure is defined as \cite{asenbeck2025}:
\begin{align}
\begin{aligned}
	\mathcal{C}_{n_1,n_2}(\rho) = \frac{1}{2} (\max_\phi \text{Tr}[O(\phi) \rho] - \min_\phi  \text{Tr}[O(\phi) \rho])
\end{aligned}
\end{align}
where $O(\phi) = \cos \phi X_{n_1,n_2} + \sin \phi Y_{n_1,n_2}$ with $X_{n_1,n_2} =\dyad{n_1}{n_2}+\text{h.c}$ and $Y_{n_1,n_2} = -\mi\dyad{n_1}{n_2}+\text{h.c}$.  Since to demonstrate that open direction we will be working with superposition states of $\ket{0}$ and $\ket{2}$, we focus on the coherence measure $\mathcal{C}_{02}$, for which the threshold is numerically obtained be $0.86$. 
Furthermore, the sensitivity of quantum non-Gaussianity (QNG) to loss is investigated via QNG coherence depth following an approach similar to Fock states in Eq.~\eqref{eq:dep}. 
%To properly accommodate coherence terms, we model the attenuation channel using a beam splitter with transmittance $T$.

\subsection*{Bunching capability of Fock states}
\begin{figure*}
\subfigure[\label{loop}] {\includegraphics[scale=0.532]{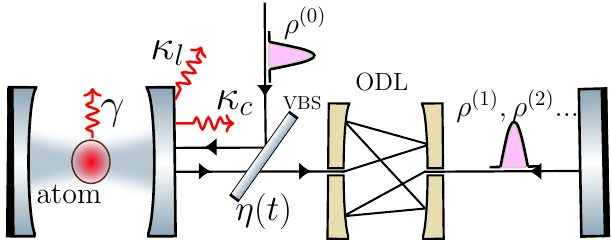}}
\hspace*{4mm}
\subfigure[\label{nol}] {\includegraphics[scale=0.532]{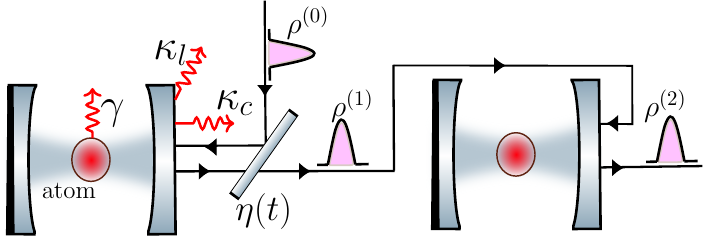}}
\hspace*{1mm}
\subfigure[\label{noi}]  {\includegraphics[scale=0.62]{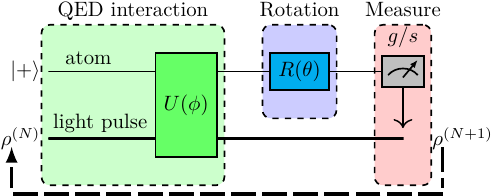}}
\caption{\subref{loop} Single cavity QED setup is used for the atom controlled phase shift optical operation \cite{Hacker_2019}, combined with a classical delay circuit \cite{yutaro_2021} to implement the iterative protocol for the high Fock states. A variable beam splitter (VBS) with time-controlled transmission $(\eta(t))$ separates the input and output pulses \cite{yutaro_2021,Yonezu-2023}. The optical delay line (ODL) is introduced by employing a advanced Herriott cell feasible with current technology \cite{Sakaguchi_2023,yutaro_2021,Yonezu-2023,Arnold_2023}.
Here, $\gamma$ and $\kappa_c$ represent the decay rates of the atom and cavity, respectively, while $\kappa_l$ denotes the scattering rate, and $g$ is the atom-cavity coupling strength.
\subref{nol}Multiple atom-cavity setup architecture
permits filtration without delay loops by implementing atomic operations following reflection processes, which is particularly crucial for waveguide-based implementations. Here, the second atom-cavity parameters are identical to the first atom-cavity setup.
\subref{noi} Circuit of a single-round of conditional Fock state filtration. The three steps of the protocol are highlighted. Each step of the iterative protocol starts with the state $\rho^{(N)}$ of the input light mode and produces the state $\rho^{(N+1)}$ of the output mode, which can then be used for the next step either in the same or subsequent cavity.
$U(\phi)$ denotes the controlled phase shift operation between light pulse (bold line) and two-level atom (thin line) in the cavity. $R(\theta)$ is unitary rotation on the two-level atom and  ($\ket{g},\ket{s}$) are the measurements on the two-level atom. 
For the Fock states, the robustness against arbitrary dephasing noise (with a jump operator, $\hat{a}^\dagger \hat{a}$) acting on light mode outside the cavity means that no phase locking in the optical delay is needed.
}\label{cir}
\end{figure*}

Further essential evaluation of filtered Fock states as the main building blocks must consider their multiple replicas.
The Fock state $\ket{n}$  can conditionally bunch with another such state and form a higher Fock state $\ket{2n}$. Such bunching phenomena are at the core of complex bosonic behaviours 
\cite{Spring_2013,Broome_2013,Tillmann_2013,Crespi_2013,Spagnolo_2014,Carolan_2014}
and are also considered in quantum technology \cite{Eaton2022,Melalkia_2023,Zheng_2023,Shunya_2024,Simon_2024}.
Fock states are typically distinguished by the number of negative annuli in the phase space, but this method is not sufficiently sensitive to contributions from higher Fock states and is prone to noise.
To overcome these issues, the Fock-state bunching capability was proposed as
a signature of the high quality of non-Gaussianity of the generated states \cite{2021_Zapl}. The bunching capability will be evaluated by the sensitive QNG hierarchy \fleq{eq:qng}. 
First, we produce a near Fock state using the filtration process as depicted in Fig.~\ref{cir}. Subsequently, $N$ copies of these states are combined in a energy-preserving  balanced linear-optics network. In order to test their potential bunching performance, we consider an extreme rare event when all the photons eventually appear in one mode. If that happens correctly, we consider all other lower bunching events to perform sufficiently well. To emulate that, all modes except the first mode are projected onto the vacuum state
\begin{align}
\rho_\text{out} =\expval {U_{N}\qty[\mathop{\otimes}\limits_{k=1}^N \rho_k] U_{N}^\dagger} {0_2~ 0_3 \dots 0_N}, 
\end{align}
where $\rho_{k}$ denotes the state in the $k_{th}$-mode. $U_N$ denotes the unitary symmetrical interference network on $N$-modes. $\ket{0_2~0_3\dots0_N}$ denotes vacuum in all modes except the first mode.
Finally, we evaluate the quantum non-Gaussian feature of the final conditional state in the optical mode $1$.

\subsection*{Sensing with Fock states and their superpositions}\label{sec:fis}
In quantum metrology, precise sensing of an unknown parameter $(\theta)$ is achieved by probing with a quantum state. 
Cramer Rao bound states that estimation error (variance) is bounded by the inverse of quantum Fisher-information ($\mathcal{F}$) i.e.
$\Delta^2 \theta \geq 1/M\mathcal{F} $ where $M$ is the number of trails.
Fock states offer a distinct advantage as optimal probes for detecting small magnitudes of phase-random optical displacement and squeezing \cite{Oh_2020}. This approach demonstrates how quantum non-Gaussianity can enhance performance, similar to findings previously analyzed for phonons \cite{podhora2022}.

A Gaussian external field can be decomposed into displacement, phase shift, and squeezing operations.
Before detailed phase-sensitive aspects of measured external fields are analysed, we often estimate first just squared amplitude $N_d$ of displacement and amount $N_s$ of the squeezing, assuming uniform distribution for phase averaging the transformation is expressed by the  map $\mathcal{M}$
\begin{align}\label{eq:scs}
\begin{aligned}
\mathcal{M} (\rho)=\dfrac{1}{4\pi^2} \int\limits_0^{2\pi} \int\limits_0^{2\pi}   d\phi_1 d\phi_2~
\hat{\mathcal{D}}(\sqrt{N_d} e^{\mi \phi_1}) \hat{\mathcal{S}}(\sqrt{N_s} e^{\mi \phi_2}) \rho \\
\hat{\mathcal{S}}^\dagger(\sqrt{N_s} e^{\mi \phi_2}) \hat{\mathcal{D}}^\dagger(\sqrt{N_d} e^{\mi \phi_1}),
\end{aligned}
\end{align} 
where $\expval{n_d} = N_d $ and $\expval{n_s} = \sinh^2 \sqrt{N_s} $ are average photon numbers  added to a vacuum state by the displacement $\hat{\mathcal{D}}(\sqrt{N_d} e^{\mi \phi_1})$ and squeezing operations $\hat{\mathcal{S}}(\sqrt{N_s} e^{\mi \phi_2})$.
Classical Fisher Information (CFI) for the ideal detection of $m$ photons is defined as 
\begin{align}\label{eq:fi}
\mathcal{F}(\theta) = \sum_m \dfrac{1}{p(m|\theta)} \qty(\pdv{p(m|\theta)}{\theta})^2
\end{align}
where $p(m|\theta) = \expval{\mathcal{M}(\rho)}{m}$ and $\theta$ represents the unknown parameter we aim to measure, such as the coherent amplitude ($N_d$) or the squeezing parameter ($N_s$). We focus on estimating either the coherent amplitude 	$(N_s=0)$ or the squeezing amplitude ($N_d =0$).
For an ideal Fock-state $\ket{m}$ \fleq{eq:fi} gives CFIs $\mathcal{F}_{D} = \frac{2 m +1} {N_d}$ and $\mathcal{F}_S = \frac{m^2+m+1} {2N_s}$ for displacement and squeezing operations.
Advantageously, in both the cases, the classical Fisher information saturates the Quantum Fisher Information (QFI) and the scheme is optimal \cite{Oh_2020}.

We employ the CFI of ideal Fock states to benchmark the phase sensing capabilities of optical states $\ket{\theta} = \cos \theta \ket{0} + \sin \theta \ket{2}$. These superpositions of Fock states are particularly suitable for phase sensing, as previously established for the motional states of ions \cite{McCormick_2019}. To validate this for our optical states, we model the sensing operation as \fleq{eq:scs} :
\begin{align} \label{eq:sem}
	\mathcal{M} (\rho) = e^{i \Theta \hat{n}} \rho e^{-i \Theta \hat{n}},
\end{align}
and  using \fleq{eq:fi} with the the simplest binary projective measurements $\dyad{\phi},I-\dyad{\phi}$ where $\ket{\phi} = \cos \phi \ket{0} + \sin \phi \ket{2}$ gives CFI for pure states as (App.~\ref{ap:fs02}):
\begin{align}
	\mathcal{F}(\Theta) = \dfrac{\sin^2(2\theta) \sin^2(2\phi) \sin^2(2\Theta)}{p(\phi,\theta,\Theta)(1-p(\phi,\theta,\Theta))},
\end{align}
where $p(\phi,\theta,\Theta)=(\cos \theta \cos \phi)^2 + (\sin \theta \sin \phi)^2 +\frac{ \sin(2\theta) \sin(2\phi) \cos(2\Theta)}{2}$.
Note that $p(\phi,\theta,\Theta)$ is symmetric with respect to $\theta$ and $\phi$ therefore, CFI as well. The maximum QFI ($\mathcal{F}=4$) is achieved when $\phi=\theta=\pi/4$.
%For a given $\theta$, the optimized CFI can be obtained using two projective measurements: $\phi=\theta$ and $\phi=\pi/4$. However, 
For superpositions limited to the states $\ket{0}$ and $\ket{1}$, the maximal QFI is only unity. Exceeding this value demonstrates enhanced phase sensitivity using higher two-photon Fock states
in the binary superposition with the vacuum.

\section{protocol}
We optimize the filtration protocol for filtering non-Gaussian states, ideally Fock states, from broadly available Gaussian states \cite{teja2023}. 
Jaynes–Cummings (JC) interactions are a natural choice for achieving controlled phase flip (CPF) between an atomic qubit and an optical mode \cite{Tiecke_2014,Hacker_2016,staunstrup2023}. Recently, dispersive interactions in superconducting circuits have also been exploited to achieve CPF \cite{deng2023}.
The protocol involves three main steps, see Fig. \ref{cir}:
\begin{enumerate}
\item QED interaction: to achieve the CPF on atom and light interacting in a cavity.
\item Rotation: unitary operation on the two levels of atom.
\item Measurement: atomic-level measurement and subsequent optical output selection.
\end{enumerate}
After the atomic measurement, the output light is used as an input state to repeat the steps using single cavity QED setup or multiple QED setup, both depicted in Fig.~\ref{loop}-\ref{nol}.

In terms of experimental implementation, the single-cavity protocol can be divided into two main components: cavity QED interaction and the classical optics control part. The classical part involves directing the pulses using a time-dependent variable beam splitter (VBS) and an optical delay line (ODL), which is necessary to ensure adequate time for atomic measurements on the atom. The optical delay can be implemented using Herriott cells or optical fibers. 

Recently the time-depended VBS are used in quantum optics experiments already \cite{yutaro_2021,Yonezu-2023}. Also ODLs have demonstrated the ability to provide microsecond-scale delays, making them ideal for cavity QED experiments with atoms \cite{Hacker_2019}. Low-loss ODLs have also been utilized in recent quantum experiments \cite{Sakaguchi_2023,yutaro_2021,Yonezu-2023,Arnold_2023}.
Alternatively, the ODL can be replaced with a quantum memory \cite{2023_Zhou, Vernaz_2018}, enabling the storage and retrieval of microsecond light pulses at desired times. 
Another promising approach for reducing delay time involves utilizing multiple atom-cavity systems visualized in Fig.~\ref{nol}, where atomic measurements are performed after the interaction. This method is particularly well-suited for artificial atoms operating on optical nanosecond timescales, such as quantum dots embedded in photonic crystal waveguides \cite{Cosacchi_2020, Najer_2019, lu2024}.

The controlled operation between the atomic states $\ket{g}$ and $\ket{s}$ and light field can be effectively written as \cite{Hacker_2019}:
\begin{subequations}\label{eq:cop}
\begin{align}
\hat{U}(\phi) &= \dyad{g} \otimes {I} + \dyad{s} \otimes \exp{\mi \phi  \hat{a}^\dagger \hat{a}},\\
e^{\mi \phi} &\approx  1 - \dfrac{2 }{1 -2 \mi (\Delta/\kappa) },
\end{align}
\end{subequations}
where $\dyad{g}+\dyad{s}  = I$. Here, $\hat{a}$ and $\kappa$ are the cavity mode annihilation operator and its decay rate. $\Delta=\omega_c -\omega_a$ is the atom-cavity detuning, where $\omega_{c}~(\omega_a)$ represent atom (cavity) frequencies, further $\Delta$ can be adjusted by changing the cavity frequency. Note that the approximation in \fleq{eq:cop} is obtained under strong atom-cavity coupling strength $(g \gg \kappa)$ and the mean frequency of input light $(\omega_L)$ is in resonance with atom $(\omega_c = \omega_L)$. The atom is prepared in the state $\ket{\psi_a} = (\ket{g}+\ket{s})/\sqrt{2}$ and the light field in a Gaussian state $\rho^{(0)}(\alpha,r,\bar{n})$ \fleq{eq:ins}. 
The circuit in Fig.~\ref{cir} will transform the input state as:
\begin{align}\label{eq:dis}
\begin{aligned}
\rho'^{(N)} &=  \hat{R}(\theta) \hat{U}(\phi) ~[\dyad{\psi_a} \otimes \rho_{ }^{(N)}]~ \hat{U}^\dagger(\phi) \hat{R}^\dagger(\theta) \\
\rho_{ }^{(N+1)} &= \dfrac{\hat{M} \rho'^{(N)} \hat{M}^\dagger} {{\text{Tr}}[\hat{M}^\dagger \hat{M} \rho'^{(N)}] } \qquad \hat{M} \in \qty{\dyad{g}, \dyad{s}},
\end{aligned}
\end{align}
where $\hat{R}(\theta)$ and $\hat{M}$ denote the two-level general unitary and atomic measurement operators. By starting with Gaussian states, such as thermal and squeezed-coherent states, and by repeating the ideal circuit, one can approach a Fock state \cite{teja2023}. All numerical simulations in this work were performed using the QuTiP \cite{2024_qutip}.

However, for the realistic physical scenario with noisy state, optical dephasing and atom-cavity scattering, the quality of Fock state preparation and its quantum non-Gaussian aspects remained unclear and challenging to predict without analysing the QNG depth.
Also, the strength of such QNG states in potential applications based on their sensitivity is unclear without analysis. Moreover, the QED circuit must be sensitively optimized to fulfil such demands under realistic conditions.

\subsection*{Thermal state and optical dephasing}
We examine the detrimental effects of photon noise and optical phase noise (decoherence) jointly within the filtration protocol. We start with the atom in $\ket{\psi_a} = {1}/{\sqrt{2}} (\ket{g}+\ket{s})$ and input light in thermal state. After atom-cavity gate the state is written as
\begin{align}
U(\phi) \qty[ \dyad{\psi_a} \otimes \rho_{\text{th}} ]U(\phi)^\dagger = &\dfrac{1}{2} \sum_{n=0}^\infty  p_n [I  +  e^{-\mi \phi n} \dyad{g}{s} \nonumber \\
&+e^{\mi \phi n} \dyad{s}{g} ] \otimes \dyad{n},
\end{align}
here $p_n = \dfrac{\bar{n}^n }{(1+\bar{n})^{n+1}}$ represents the initial thermal Bose-Einstein statistics of a single-mode light field. Further performing atomic state rotation $R(\theta) = \sigma_z - \sigma_x $ and measuring the atom projects the optical field to 
\begin{subequations}\label{eq:thm}
\begin{align}
\rho = 
& {N_g} \sum_{n=0}^\infty c_n  \cos^2\qty(\dfrac{n\phi}{2}) \dyad{n} \quad \text{for} \quad \ket{g},\\
\rho =  & {N_s} \sum_{n=0}^\infty c_n  \sin^2\qty(\dfrac{n\phi}{2}) \dyad{n} \quad \text{for} \quad \ket{s},
\end{align}
\end{subequations}
here $N_{\qty{g,s}}(\phi)$ is the normalization constant. By adjusting $\Delta/\kappa$ to change the controlled phase $\phi$, we can asymptotically reach any desired Fock-state with high fidelity. It's important to note that \fleq{eq:thm} always result in a diagonal matrix (in Fock basis). Therefore, by repeating the above steps  we can sequentially remove Fock numbers to filter a given Fock state without locking phase of light pulse outside the cavity.
As shown above, it's evident that for the Fock filtration process, only the diagonal terms of the input light are relevant. Hence the filtration will be immune to dephasing of the form:
$
\mathcal{D} (\rho) = \hat{L} \rho \hat{L}^\dagger - \dfrac{1}{2} \hat{L}^\dagger \hat{L} \rho - \dfrac{1}{2} \rho \hat{L}^\dagger \hat{L},
$
where the jump operator $L = \hat{a}^\dagger \hat{a}$. Although dephasing in the light pulse outside the cavity does not affect the filtration of Fock states, dephasing during the atom-photon interaction will impact the process (see Appendix).
Identical results to those in Fig.~\ref{inp} and Table.~\ref{tab:mes} are obtained even with optical dephasing noise.
It is advantageous for future experimental tests of the Fock states as the optical field does not have to be phase-locked if direct photo-number detection or multiplexed single-photon detectors are used to measure photon number statistics.
%and no optical phase locking is required between the steps of protocol in Fig.\ref{cir}.

\subsection*{Atom-cavity loss}
The  dynamics of  atom-field interactions are best captured using virtual cavities \cite{2019_klaus,2020_kii}, however this approach makes it very difficult to numerically solve for higher Fock states. Hence, we will use this approach only when the dimensions of the density matrix in the Fock basis are manageable.
Another approach is to account for noise using Kraus operators \cite{2022_hastrup}. 
This method is suitable for steady-state dynamics and is effective for simulating higher Fock states.
Since we will be using high Fock states, we will adopt the Kraus operator approach.
The controlled operation between atom and light depend on two important metrics \cite{2022_hastrup} , cooperativity $C$ and cavity efficiency $\beta$:
\begin{align}\label{eq:ck}
C=\dfrac{g^2}{\kappa \gamma}, \quad \beta = \dfrac{\kappa_c}{\kappa}. 
\end{align}
Here, $g$ and $\gamma$ are the atom-cavity coupling strength and the atomic decay rate outside the cavity, respectively. The total cavity decay rate $\kappa = \kappa_c + \kappa_l$ is split into cavity mode decay and scattering losses ($\kappa_l$). The efficiency $\beta$ denotes the ratio of light emission into cavity over total emission.  In all our simulations, unless stated otherwise, we use the numbers targeted by current technology: $C=250$ and $\beta =0.99$. 

In conventional cavities, cooperativity on the order of $C \sim 10-100$ can be achieved by increasing the finesse more than $10^5$ \cite{reiser_2015}, and $C \sim 10^4$ can be achieved by coupling atoms to photonic crystal cavities \cite{2018_chang}. In conventional Fabry-Pérot cavities, the typical values are around $\beta \approx 0.92$ \cite{Hacker_2019}, while $\beta$ factors close to unity are demonstrated in semiconductor cavities \cite{Najer_2019} and photonic crystal waveguides \cite{staunstrup2023}.
Furthermore, the ODL can also result in photon losses that effectively can be viewed as increasing the scattering losses $\kappa_l$, which reduces both the cooperativity $C$ and the cavity efficiency $\beta$, as discussed in the Appendix.~\ref{ap:srno}.

\section{Results}
\subsection*{Filtering Fock states and their superpositions from Gaussian states}
\begin{figure*}[!t]
\hspace{-1cm}
\subfigure[\label{ph_dist}] {\includegraphics[width=17cm,height=6.5cm]{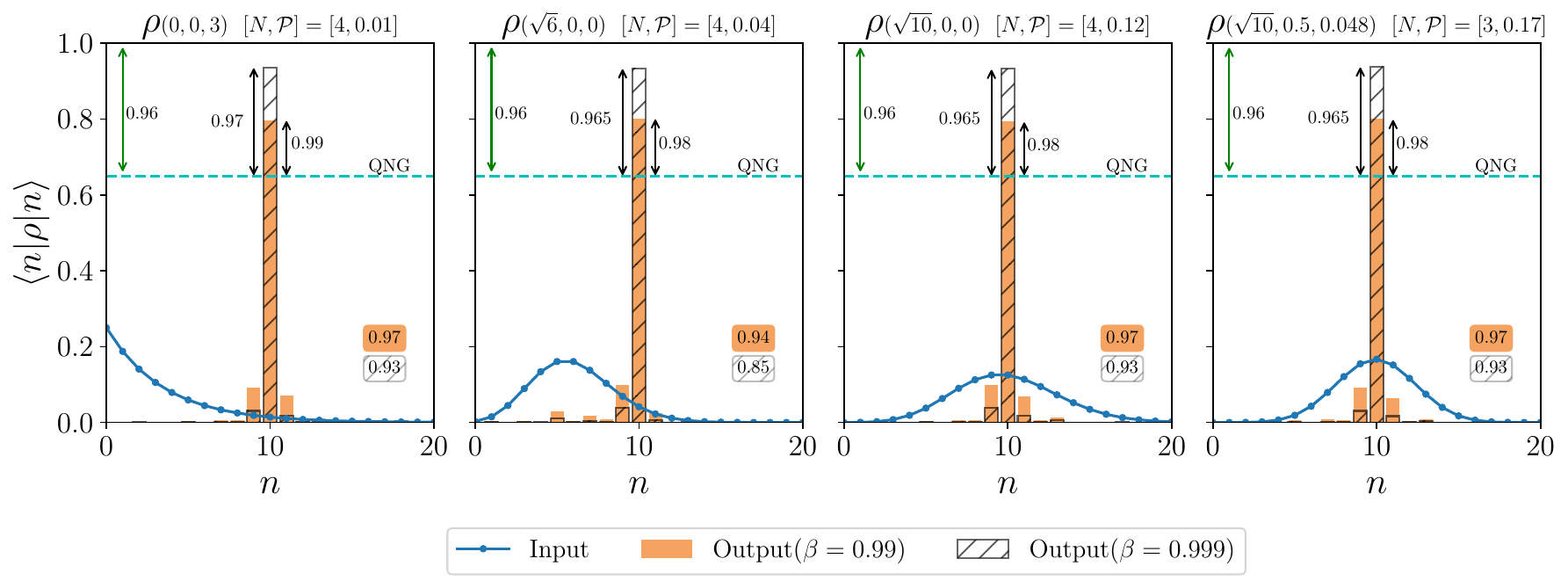}}
\subfigure[\label{outs}] {\includegraphics[scale=0.4]{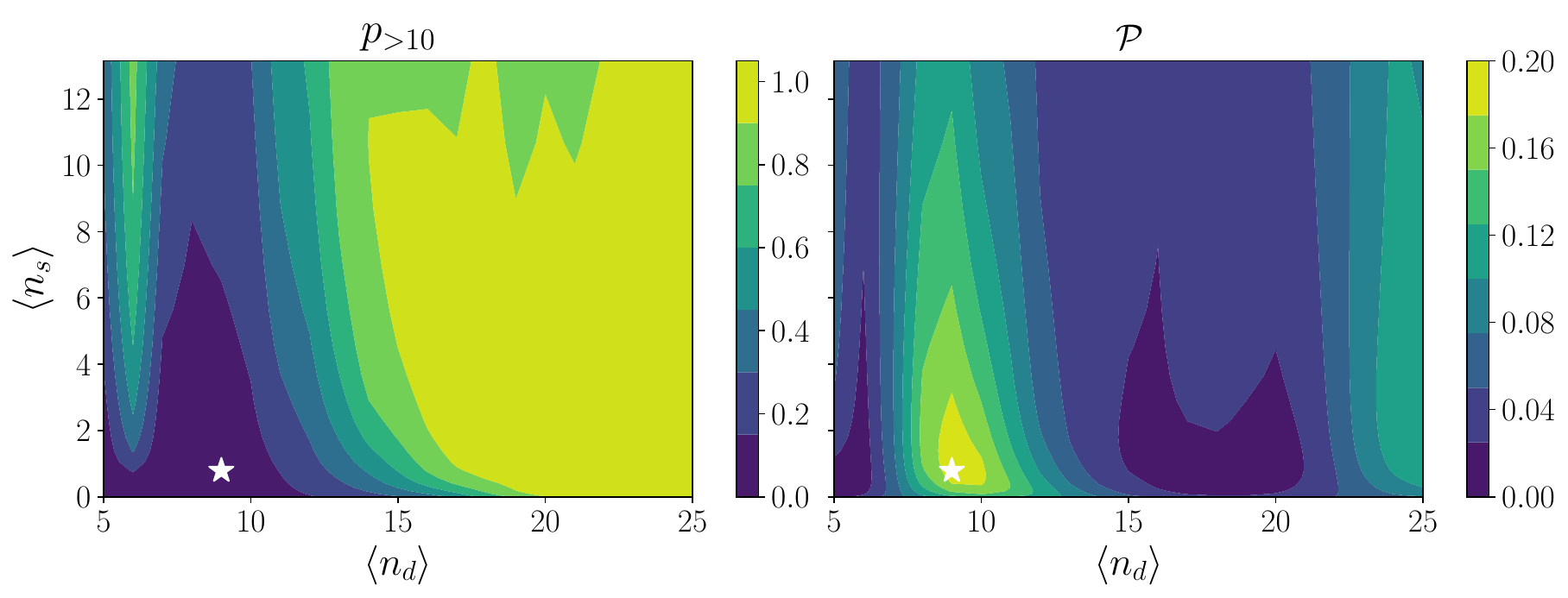} }
\hspace{-0.4cm}
\subfigure[\label{tmsv}]{ {\includegraphics[width=5.2cm,height=4.5cm]{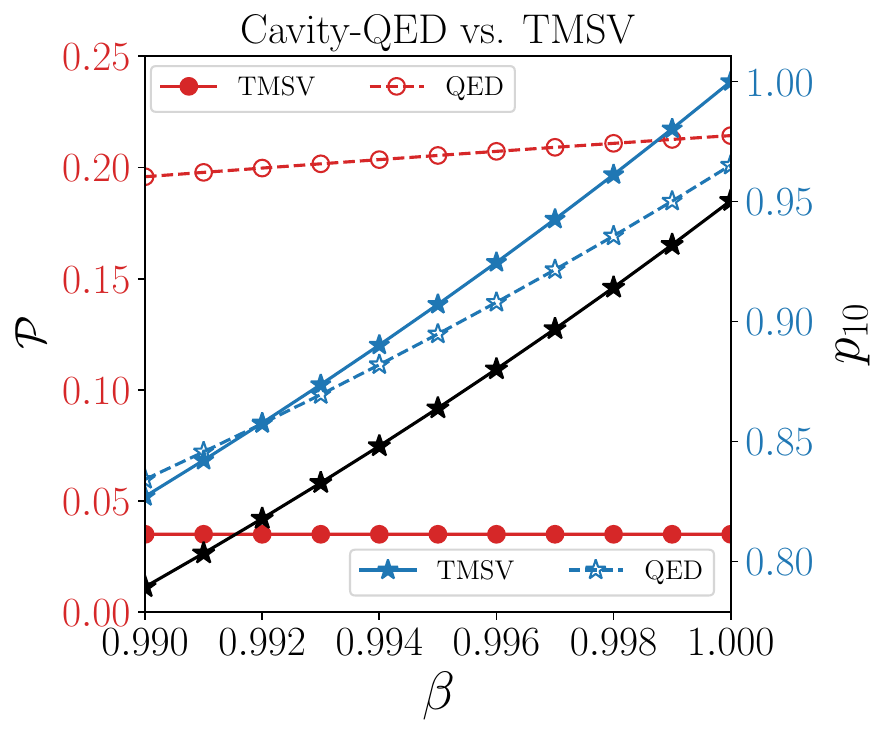} }}
\caption{\subref{ph_dist}
Filtration of Fock state $\ket{10}$ from various Gaussian sources (\fleq{eq:ins}) for different cavity noises ($\beta$ \fleq{eq:ck}). 
The input states are denoted as $\rho(\alpha,r,\bar{n})$, as defined in \fleq{eq:ins}, with the blue lines representing the photon distribution of these input states. The required number of access to the cavity and the success rate of atomic measurements is denoted by $[N,\mathcal{P}]$. 
The QNG line represents the threshold  $\mathcal{T}_{10}$ (\fleq{eq:qng}). The arrows denote quantum depth for the filtered Fock states and ideal state ($\ket{10}$) before they drop below the QNG threshold. The lower numbers in boxes show the noise  of the Fock state before they drop below the r-QNG criterion.
\subref{outs} shows the higher Fock state contributions (left) and atomic measurement success ($\mathcal{P}$) as a function of average photon numbers corresponding to displacement and squeezing in the input optical state parametrized as $\expval{n_d}=\abs{\alpha}^2$ and $\expval{n_s}= \sinh^2 r$.
The marked point ($\star$) represents the squeezed coherent state $\rho(\sqrt{9},0.8,0)$ which is compared with an example of two-mode squeezed vacuum (TMSV) in \subref{tmsv}.
\subref{tmsv} Comparison of cavity-QED filtration and TMSV protocol for Fock-state generation.
$\mathcal{P}$ denotes success probability and $p_{10}$ denotes fidelity.
All TMSV plots are simulated with $\beta = 0.99$ and unit efficiency for PNRDs (see Sec.~\ref{ap:tmsv}), except the black-line which shows the result for PNRDs with efficiency of $0.995$. The cavity-QED parameters in Figs.~\subref{outs} and \subref{tmsv} are set to $\qty{C,\beta} = \qty{250,0.99}$.
}\label{inp}
\end{figure*}

The input Gaussian states are generally represented by a displaced squeezed thermal noise:
\begin{align}\label{eq:ins}
\begin{aligned}
\rho(\alpha,r,\bar{n}) =   \hat{\mathcal{D}}(\alpha)  \hat{\mathcal{S}}(r)  ~ \rho^{\text{th}}_a(\bar{n}) ~ \hat{\mathcal{S}}^\dagger(r)  \hat{\mathcal{D}}^\dagger(\alpha), 
\end{aligned}
\end{align}
where $\rho_{\text{th}}=\frac{1}{1+\bar{n}} \sum_{n=0}^\infty \qty(\frac{\bar{n}}{1+\bar{n}})^n \dyad{n}$ denotes the thermal state with a mean photon number of $\bar{n}$.
$\hat{\mathcal{D}}(\alpha)$ and 
$\hat{\mathcal{S}}(r) $ are the displacement and squeezing operators, as before.

Fig.~\ref{inp} demonstrates how different Gaussian input states produce targeted Fock states within a maximum of $N=4$ protocol rounds. Notably, even states with large photon noise like thermal states efficiently yield Fock states. However, the probability of success remains limited to $\mathcal{P} = 0.01$.
By optimization, as shown in Fig.~\ref{inp} and Table~\ref{tab:mes}, the success probability of atomic measurements can reach upto $(\mathcal{P} = 0.17)$ using just Gaussian input states. 
Already, Poissonian distribution of coherent states increases the success rate by one order ($\mathcal{P}> 0.1$) when optimized to have a probability maximum at the desired Fock-state $\ket{10}$, as visible in Figs.~\ref{ph_dist}. Further improvement is possible by Gaussian squeezing concentrating the photon number distribution more around that maximum. The maximum success probability ($\mathcal{P} \approx 0.20$) can be reached when the Gaussian squeezing is optimally adjusted to the displacement of the coherent state, see Fig.~\ref{outs}.  
It demonstrates the power of the emitters in cavities with the delay elements to be used as filters rather than heralding detectors.

To use an advantage of a loss-tolerant r-QNG hierarchy, it is crucial to minimize contributions from higher Fock states. For instance, when filtering the state $\ket{10}$, the probability $p_{\scriptsize{>10}} = \sum_{n=11}^\infty\ev{\rho}{n}$ should be minimal. Higher Fock-state contributions can be reduced by lowering the mean of the input light, as shown with the states $\rho(0,0,3)$ and $\rho(\sqrt{6},0,0)$. However, this reduction comes at the cost of the success rate of atomic measurements, resulting in $\mathcal{P} < 0.05$. By allowing for small contributions from higher Fock states, the success rate can be improved by an order of magnitude, achieving $\mathcal{P} > 0.12$, as illustrated with the states $\rho(\sqrt{10},0,0)$ and $\rho(\sqrt{10},0.5,0.048)$. This trade-off is beneficial for experiments that need to tolerate more significant losses.
 
In Fig.~\ref{outs}, we show the contribution from higher Fock states $(p_{\scriptsize{>10}})$ and atomic success rate $(\mathcal{P})$ as a function of photon numbers corresponding to displacement ($ \expval{n_d}= \abs{\alpha}^2 $) and squeezing  ($ \expval{n_s}= \sinh^2 r $) , and it is clear that the optimized range to distil the state $\ket{10}$ is $\expval{n_d}\approx 10$ and $\expval{n_s} < 2$.
It is advantageous, therefore, to use coherent states with $\langle n_s\rangle \sim 0$ from suitable lasers tunable to the atomic transition for the first tests of such iterative protocols.

In   Fig.~\ref{tmsv}, we compare the filtration protocol using the squeezed coherent state $\rho(\sqrt{9},0.8,0)$ with the state-of-the-art Fock state generation utilizing ideal TMSV and ideal PNRDs \cite{Tiedau_2019,Lachman_2019,Cooper_2013,Yukawa_2013}. We chose the TMSV with the maximum probability for the state $\ket{10,10}$ i.e. $\lambda=\tanh(1.85)$, see Appendix.~\ref{ap:tmsv}.
The success probability (red lines) of the QED filtration protocol ($\mathcal{P} \approx 0.20$) is six-times higher than that of the ideal TMSV and PNRD protocol ($ \mathcal{P} \sim 0.03$). Although the fidelities ($p_{10}$) of ideal TMSV with ideal PNRDs (blue lines) are slightly better than those of the filtration protocol, the addition of already slight inefficiency of PNRD ($\beta = 0.995$) \cite{Provazn_2020} reduces the fidelities below those achieved by the QED filtration protocol (black line). Overall, this demonstrates that the QED filtration protocol offers  higher success rate ($\sim 20\%$) without sacrificing the quality of generated Fock-states. This improvement without compromise must be confirmed in the experiments, which can stimulate further developments of the cavity-delay techniques.

\begin{figure}
	\subfigure[\label{st_02}]{\includegraphics[scale=0.415]{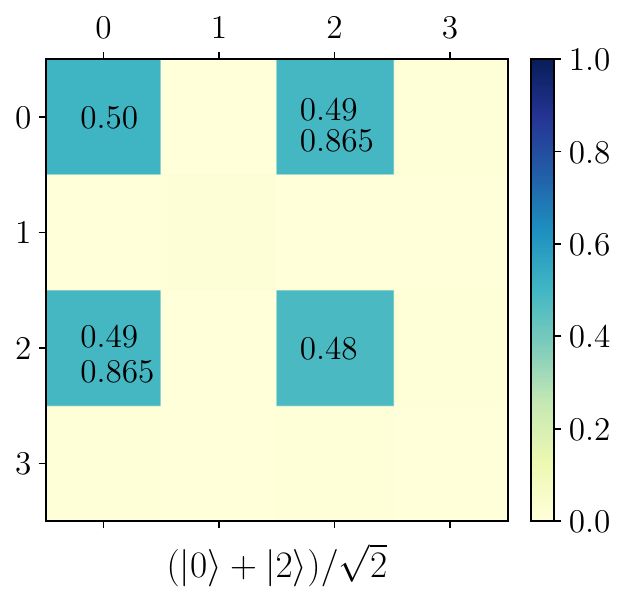}}
	\subfigure[\label{c_02}]{\includegraphics[scale=0.40]{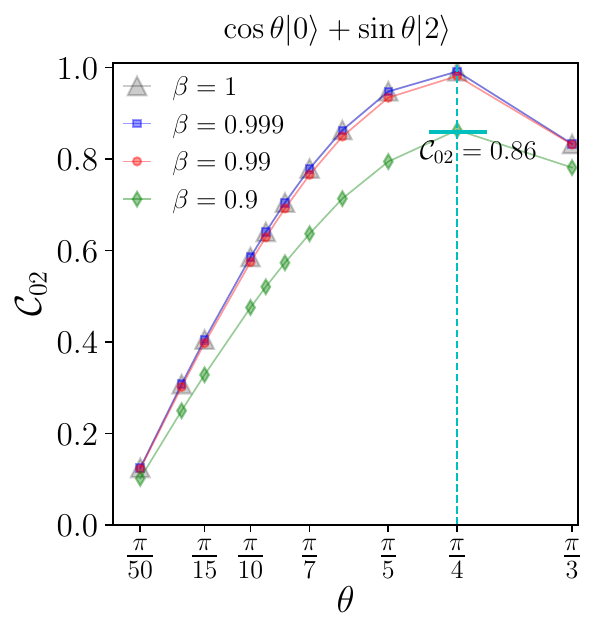}}
	\caption{\subref{st_02} Density matrix for the filtered superposition of the Fock states $(\ket{0} + \ket{2})/\sqrt{2}$. Optimized displaced squeezed vacuum state $\rho(1.035,0.247,0)$ is used as input.  Cavity parameters: $\qty{C,\beta} = \qty{205,0.99}$ yields an atomic success probability of $0.515$ and depth of QNG coherence of 0.865. \subref{c_02} Coherence measure of filtered states $\cos\theta\ket{0} + \sin\theta\ket{2}$, with the threshold ($\mathcal{C}_{02}=0.86$, green-dash) indicating genuine QNG coherence for $\theta=\pi/4$ \cite{kov2024}. All states are generated via single atom-cavity interaction. Already $\beta=0.99$ is sufficient to approach ideal performance $\beta=1$.} \label{fg:cfs02}
\end{figure}
The filtration protocol repeatedly exploits CPF in sequence 1-2-3 to filter out a Fock state from a given state and, therefore, it might look it lacks flexibility in preparing a binary superposition of the Fock states. However, we show that by optimizing the input state, the light filtration technique allows for the preparation of states of the type $\cos \theta \ket{0} + \sin \theta \ket{2}$.
It is an essential example that this method is capable of higher and ultimate QNG coherence, success rate and is
more variable for the optical state preparation than was considered in Ref. \cite{teja2023}.

Displaced squeezed vacuum states \fleq{eq:ins} with low coherent and squeezing amplitudes ($\alpha,r$) can be approximated as (see Appendix.~\ref{ap:scs})
\begin{subequations}
	\begin{align}
		\psi_\text{sc} \approx  C_0 \qty[ \ket{0}  + C_1\ket{1}+  C_2\ket{2} +  C_3 \ket{3} ],\\
		\sum_{k=0}^3 \abs{C_k}^2 \sim 1, \label{eq:sm}
	\end{align}
\end{subequations}
after atomic rotation and measurement, projects the light state to
\begin{subequations}
	\begin{align}
		\ket{\theta} \approx &   \cos \theta \ket{0}  + \sin \theta \ket{2} ,\\
		\theta \approx &\tan^{-1}\qty({ \dfrac{\tanh r}{2\sqrt{2}}  H_2\qty[\dfrac{\alpha e^{r} }{\sqrt{ \sinh{2r}}}]}). \label{eq:tht}
	\end{align}
\end{subequations}
For a given $\theta$, we numerically optimize $\alpha$ and $r$ such that \eqref{eq:sm} and \eqref{eq:tht} are optimized, but the figures are computed without any truncation. 
For example to filter the state $(\ket{0}+\ket{2})/\sqrt{2}$, the optimized input $\rho(1.035,0.247,0)$ gives $\theta= \pi/4$ and $\sum_{k=0}^3 \abs{C_k}^2 = 0.935$
with coherence ${\cal C}_{02}$ represented by the off-diagonal elements over the ultimate QNG threshold 0.86 and close to maximal value, as visible in Fig.~\ref{c_02}.

Our results, shown in Fig.~\ref{fg:cfs02}, demonstrate that cavity QED filtered states $(\ket{0}+\ket{2})/\sqrt{2}$ achieve high quality and surpass the absolute QNG coherence threshold, even with substantial cavity losses ($\beta = 0.9$). Furthermore, the coherence QNG-depth with $\beta=0.99$ reaches $T=0.875$,
and for $\beta=0.999$ even $T=0.866$, close to the ideal cavity ($\beta=1$) depth of $T=0.865$.
Therefore, for $\beta=0.99$ we already reach 92.5\% of the maximal loss tolerance for QNG coherence.

Beyond generating Fock states, this establishes cavity QED as an effective platform for producing superpositions of Fock states with unprecedented quality using only single atomic measurements. Such balanced superpositions $(\ket{0}+\ket{2})/\sqrt{2}$ surpass cat-state preparation capabilities \cite{2019_Daiss} and open new avenues for advancing optical sensing protocols, building on mechanical implementations already demonstrated \cite{McCormick_2019}.
Note that in Fig.~\ref{c_02}, $\theta = \pi/2$ is excluded since achieving high-quality state $\ket{2}$ requires two reflections.

%%%%%%%%%%%%%%%%%%%%%%%%%%%%
\subsection*{QNG-depth and bunching capability of high Fock-states}
\begin{figure*}
	\subfigure[\label{td}] {\includegraphics[scale=0.46]{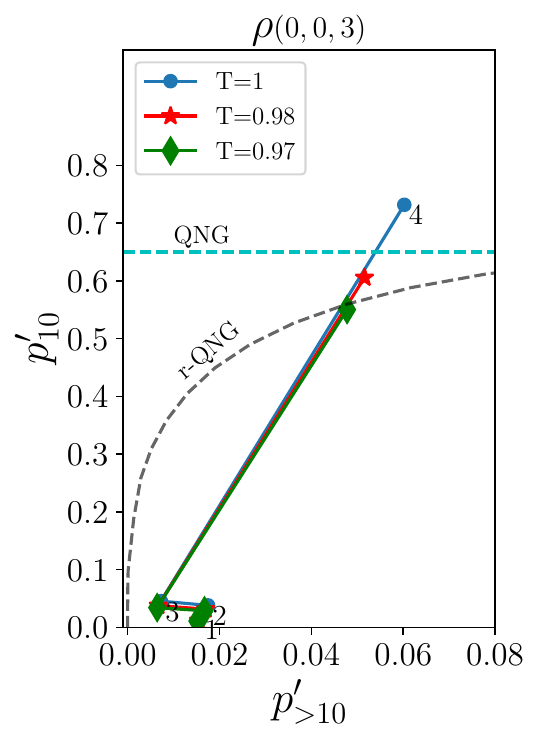}}\hspace{-0.23cm}
	\subfigure[\label{chdd}] {\includegraphics[scale=0.46]{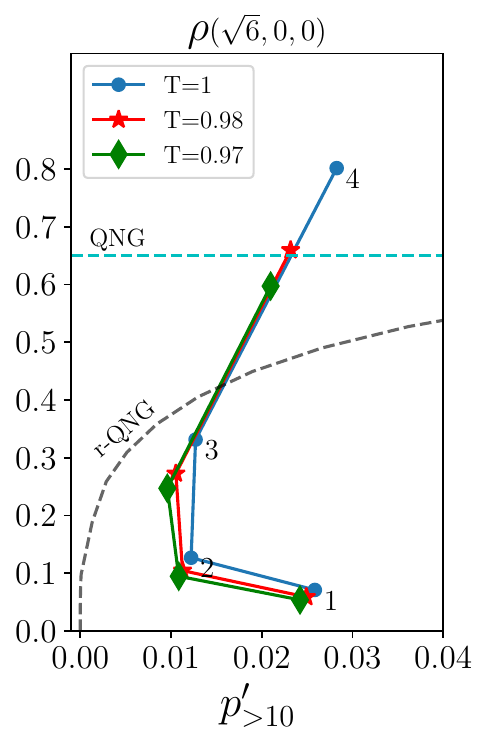}} \hspace{-0.25cm}
	\subfigure[\label{chd}] {\includegraphics[scale=0.46]{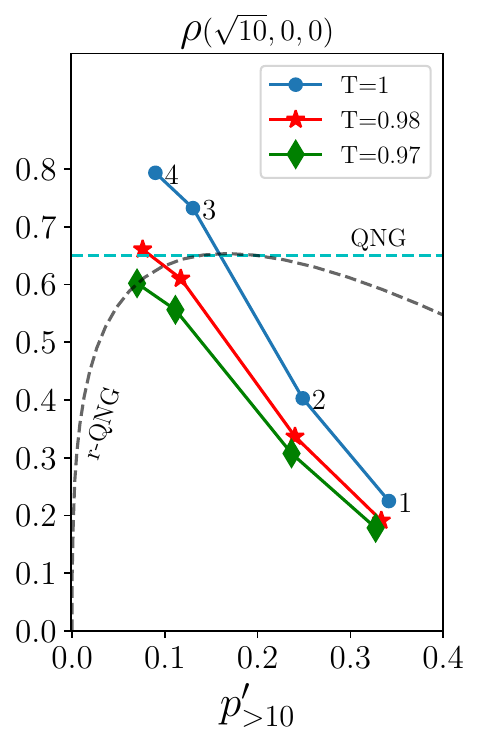}} \hspace{-0.25cm}
	\subfigure[\label{sqd}] {\includegraphics[scale=0.46]{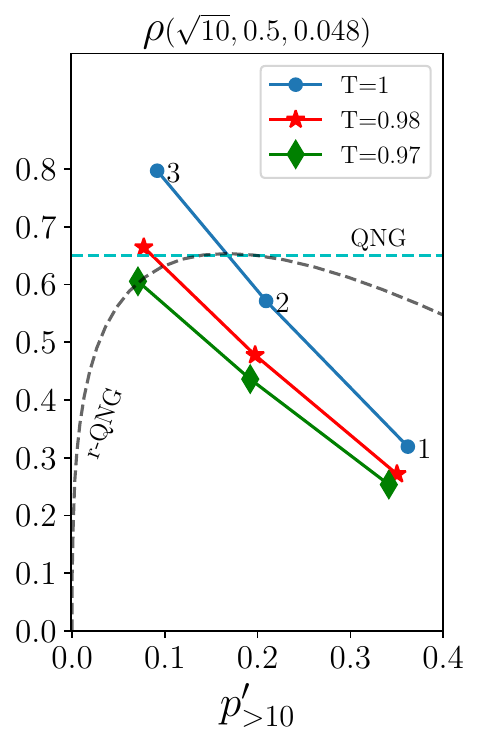}}\hspace{-0.25cm}	
	\subfigure[\label{bs0}] {\includegraphics[scale=0.33]{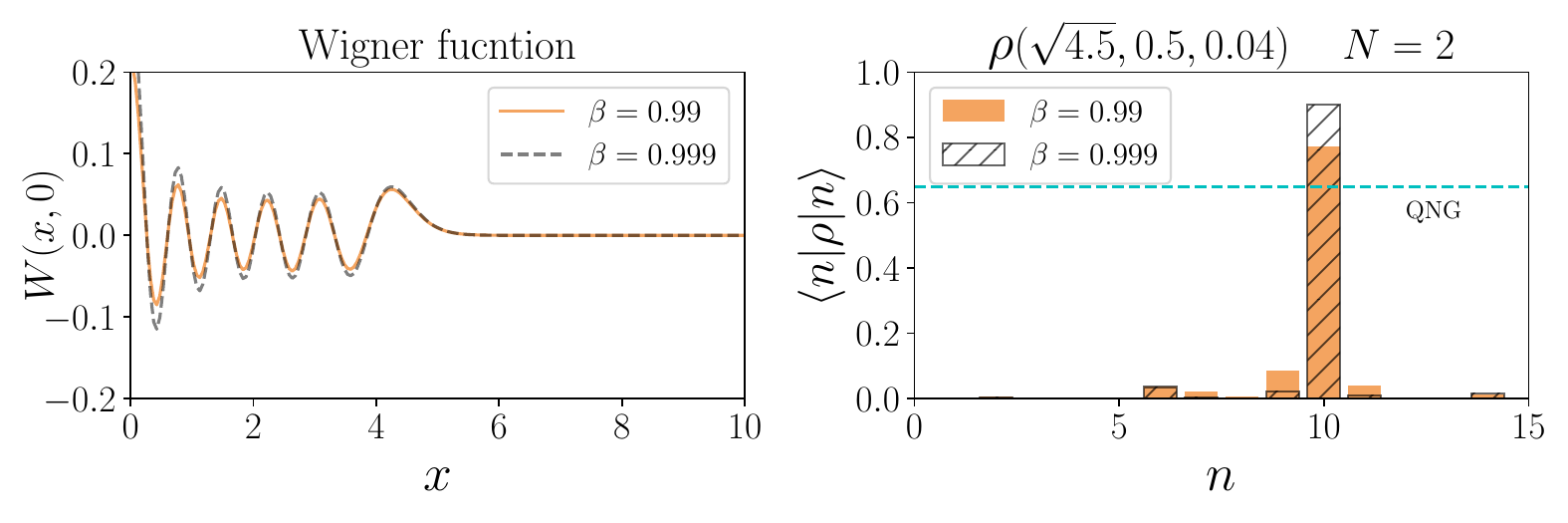}}
	\subfigure[\label{bs6}] {\includegraphics[scale=0.33]{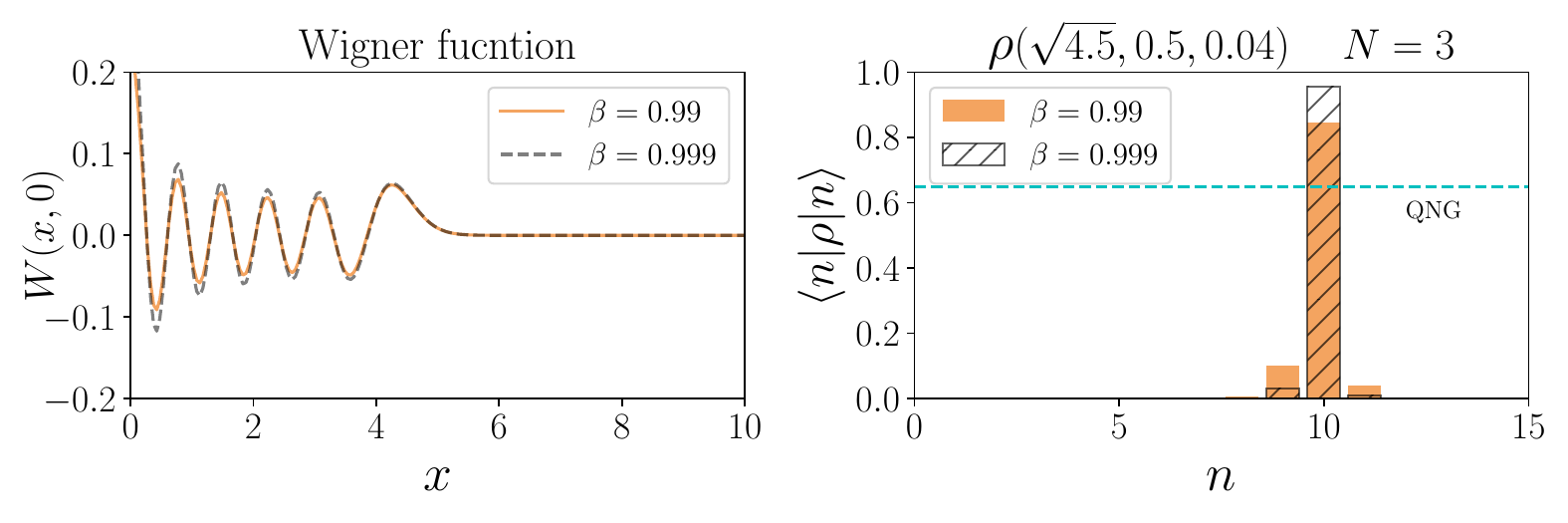}}
	\caption{ \subref{td}-\subref{sqd} QNG-depth of the Fock-states of light. Top labels show the various states used in filtration. The numbers $1$-$4$ denote the ordering of atom-cavity reflections. The probabilities $p'_{10}$ and $p'_{>10}$ are calculated using \fleq{eq:dep}, for each state after atom-cavity reflection. $T$ is the transmittance of the output lossy channel. The r-QNG (black dashed curve) represents the relative criterion, while the dashed line represents the absolute criterion ($\mathcal{T}_{10}$), \fleq{eq:rqng}. The cavity parameters are set to $\qty{C,\beta} = \qty{250,0.99}$, see description in the text.
	%\end{figure*}
	%\begin{figure*}	
	\subref{bs0}-\subref{bs6} Wigner distribution and Fock-state bunching capability.
		Fock-state $\ket{5}$  filtered from the state $ \rho(\sqrt{4.5},0.5,0.04) $ 
		is bunched in two modes to generate $\ket{10}$. 
		Figs.~\subref{bs0} and \subref{bs6} shows the state after two and three atomic measurements. QNG denotes the absolute threshold (\fleq{eq:qng}) for the Fock-state $\ket{10}$.  The cavity parameters are set to $\qty{C,\beta} = \qty{250,0.99}$.
	}\label{fig:dep} 
\end{figure*}

To understand the distance of the predicted bars in Fig.~\ref{ph_dist} from the quantum non-Gaussianity threshold, we examine in detail the depth of the approximate high Fock-states obtained from the filtration protocol. We calculate $p'_{10}$  (\fleq{eq:dep}) for states after each atom-cavity reflection (plus measurement), and then compare it with  $\mathcal{T}_{10}$.
In Fig.~\ref{ph_dist}, we observe that Fock-states filtered with \(\beta = 0.999\) from various Gaussian inputs exhibit a QNG-depth ranging from \(T = 0.97\) to \(0.965\). For comparison, lowering \(\beta\) to \(0.99\) reduces the QNG-depth to \(T = 0.98\). As a reference point, the ideal Fock-state \(\ket{10}\) has a QNG-depth of \(T = 0.96\) (indicated by the green arrow). Therefore, even with practical scattering losses (\(\beta = 0.999\) and \(\beta = 0.99\)), we achieve 75\% and 50\% of the ideal QNG depth, respectively.

In Fig.~\ref{fig:dep} we show the QNG and r-QNG depths of filtration protocol after each atom-cavity reflection (marked by numbers). 
The dashed lines show the QNG criterion obtained by numerically solving \fleq{eq:qng} for the state $\ket{10}$.
From  Figs.~\ref{chdd}-\ref{chd} we see that three atomic measurements are required for coherent states to surpass the threshold in the hierarchy \fleq{eq:qng} while the thermal state requires four atomic measurements (Fig.~\ref{td}). 
Fig.~\ref{sqd} shows that by optimizing the squeezing and displacement parameters, three atomic measurements are sufficient for the state $\rho(\sqrt{10}, 0.5, 0.048)$ to surpass the QNG threshold.
Also note that the states after the first filtration-step (marked 1) are experimentally already demonstrated Schr\"odinger cat states formed by a superposition of phase randomized Gaussian states \cite{Hacker_2019,Li_2024}. Although they are non-Gaussian states, due to contributions from higher photon numbers prevents them from surpassing even the relative QNG criterion for the Fock-states.
More filtration steps are needed to reach the r-QNG threshold and eventually the absolute one. However, a positive step can be visible after the first delay and second reflection. During the process, the multi-photon contributions $p_{>10}$ can be compromised with the probability of success $\cal{P}$.

In Fig.~\ref{fig:dep}, although the filtered Fock-states are similar, the QNG and r-QNG depths are highly sensitive to the photon distribution. To achieve a high depth, it is crucial to have high $p_{10}$ probabilities and low contributions from higher Fock-states $p_{>10}$.
While QNG and r-QNG depths characterize the noise tolerance of Fock-states obtained through filtration, the total efficiency of the process is determined by the atomic measurements.
As shown in Fig.~\ref{outs} and Table~\ref{tab:mes} in the Appendix, there is a trade-off between low contributions from higher Fock-states and the success rate of atomic measurements. Similarly there is an inverse correlation between atomic measurements and QNG-depth; for example, in Fig.~\ref{chd} the r-QNG depth with $T=0.97$ using 4 measurements is identical to the depth with $T=0.98$ using 3 measurements.

The quality of generated Fock-states is not only quantified by their robustness to loss and, moreover, also their possibility of combining with resources similar to more capable ones.
In Figs.~\ref{bs0}-\ref{bs6}, we show the bunching capability \cite{2021_Zapl} of the Fock-state with five photons obtained from the a squeezed coherent state after two and three atomic measurements. 
We simultaneously use continuous-variable interference effects in the Wigner function exhibiting adequate negative regions and discrete variable characteristics in photon number distribution. Both confirm the high quality of the bunching capability of two copies of five photon state. It allows them to be used for advanced bosonic applications, like gates on binomial quantum error correction codes \cite{2016_mich}.

\section{sensing}%Sensing using Fock-states of light}
\begin{figure*}
	\subfigure[\label{fc6}] {\includegraphics[scale=0.38]{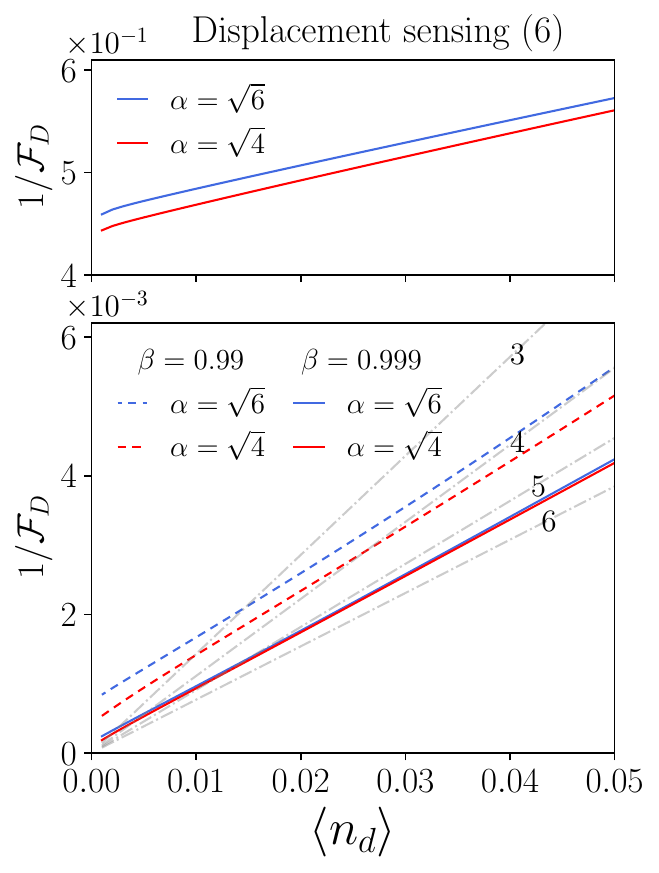}} 
	\subfigure[\label{fr6}] {\includegraphics[scale=0.38]{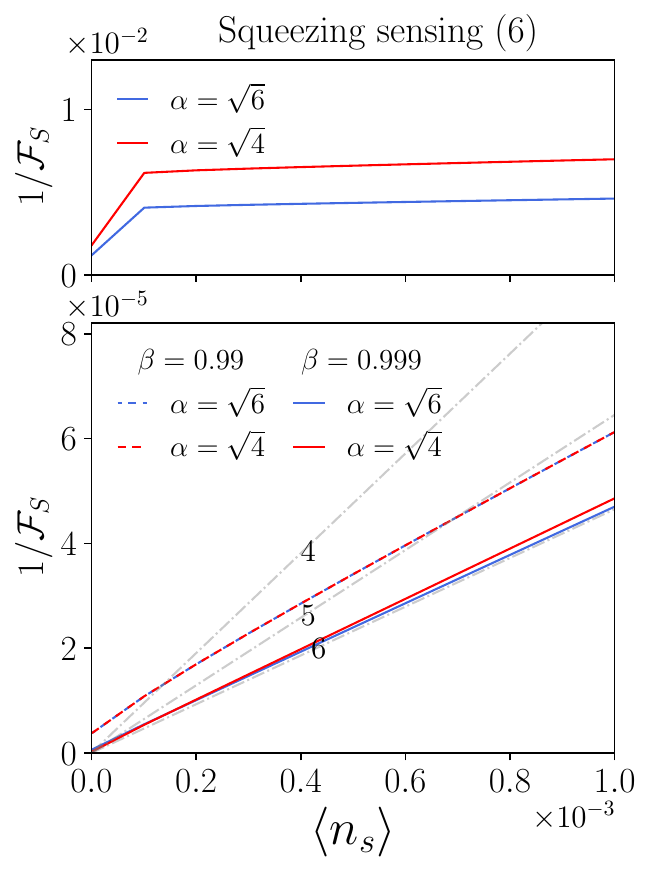}}
	\subfigure[\label{fc}] {\includegraphics[scale=0.38]{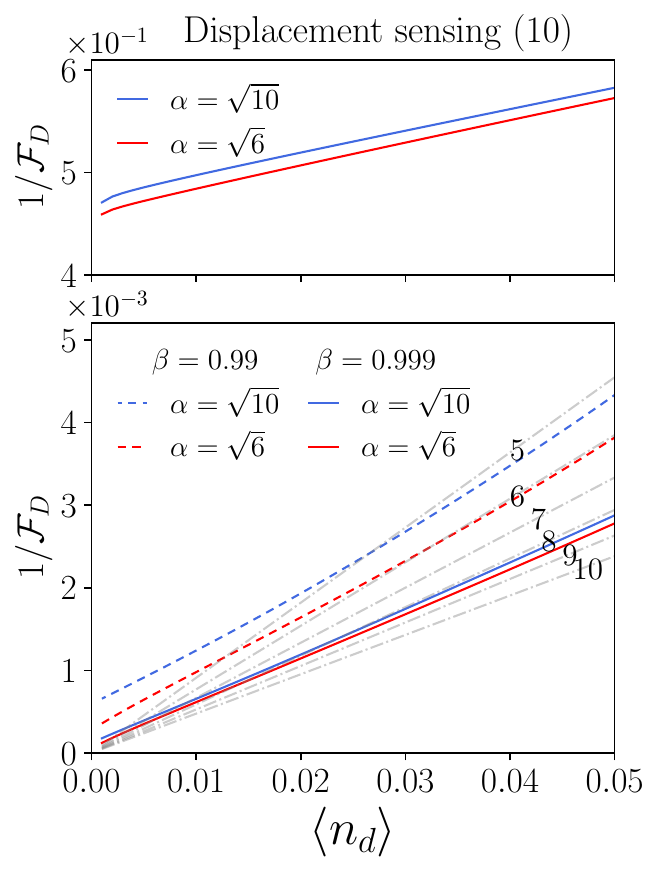}} 
	\subfigure[\label{fr}] {\includegraphics[scale=0.38]{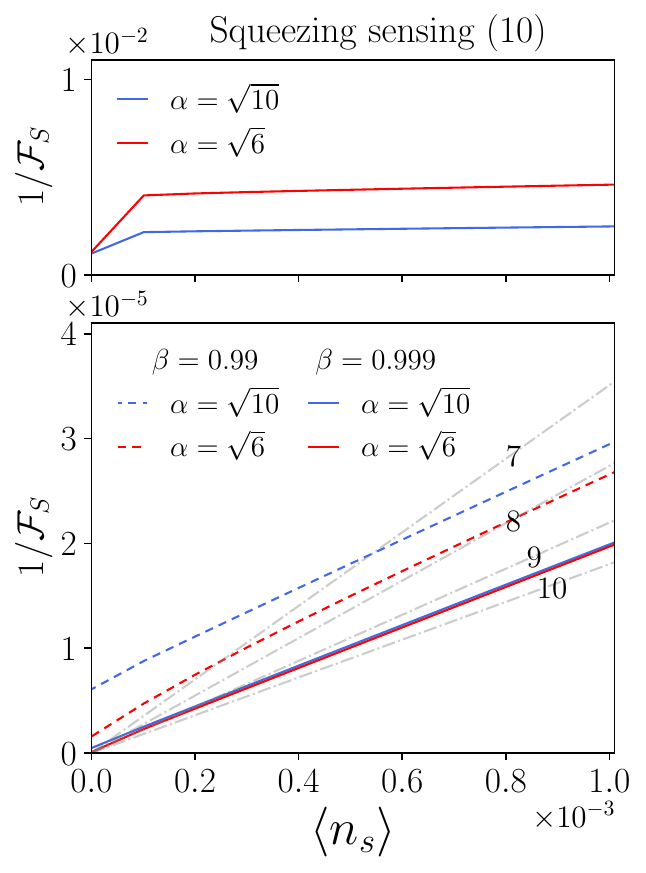}}
	\caption{Sensing photon numbers corresponding to displacement (\subref{fc6}, \subref{fc}) and squeezing (\subref{fr6}, \subref{fr}) using filtered states derived from coherent states is presented. The numbers $(10, 6)$ in the top labels indicate the targeted Fock states.
		The average photon numbers are given by $\expval{n_d} = N_d$ and $\expval{n_s} = \sinh^2 \sqrt{N_s}$, while $\mathcal{F}_D$ and $\mathcal{F}_S$ denote the Fisher information related to displacement and squeezing operations, respectively (see Sec.~\ref{sec:fis}).
		The top plots depict the estimation using the input coherent states with $\alpha = \qty{\sqrt{4},\sqrt{6}, \sqrt{10}}$, while the bottom plots illustrate the estimation of the output filtered states from the corresponding inputs for various $\beta$ (\fleq{eq:ck}) values. Estimations for ideal Fock states ($\ket{5}$–$\ket{10}$) are shown by the dot-dashed (black) lines. The cooperativity for the QED filtration is set to $C = 250$.}\label{fg:fff}
\end{figure*}
To perform further analysis we advantageously use that the Fock-states of light are already resources for the advanced quantum sensing of the amplitude of the external displacement (and squeezing) \cite{Oh_2020,Han_2023}. We analyse an estimation error to verify if quantum non-Gaussian states provide its reduction conclusively.

In Fig.~\ref{fg:fff}, we numerically calculate the Fisher information \fleq{eq:fi}, with $\rho$ being the state obtained from the cavity-QED filtration. Although the statistical properties of the filtered states are close to the Fock states $\ket{6}$ and $\ket{10}$ (Fig.~\ref{ph_dist}), their sensing capabilities differ. When sensing small values of $\expval{n_d}$ and $\expval{n_s}$, noise introduces an offset from the thresholds determined by the ideal Fock states \cite{podhora2022}. Interestingly, squeezing values exhibit better performance than displacement by a factor of $100$. Furthermore, improving $\beta$ not only minimizes the offset but also enhances sensing capabilities.

In Fig.~\ref{fc6}, displacement sensing with the targeted state $\ket{6}$ surpasses the performance of the ideal Fock state $\ket{5}$ for $\beta=0.999$, after an initial offset where the Fisher information drops below that of lower Fock states. For lower $\beta=0.99$, the Fisher information from the generated states surpasses that of the ideal Fock state $\ket{3}$, emphasizing the sensitivity of Fisher information to $\beta$. A similar trend is observed in Fig.~\ref{fr6} for squeezing sensing, where for $\beta=0.99$, the performance exceeds the threshold of Fock state $\ket{4}$, and for $\beta=0.999$, it approaches the performance of the ideal Fock state $\ket{6}$.  

In Fig.~\ref{fc}, targeting the Fock state $\ket{10}$ shows that Fisher information for the generated states surpasses that of the ideal states $\ket{5}$ for $\beta=0.99$ and $\ket{8}$ for $\beta=0.999$, after an initial offset. Similarly, in squeezing sensing shown in Fig.~\ref{fr}, the performance surpasses the Fock state $\ket{7}$ for $\beta=0.99$ and $\ket{9}$ for $\beta=0.999$.  

These results demonstrate that targeting higher Fock states provably improves the sensing protocol. Moreover, lowering the initial coherent state from $\alpha=10$ to $\alpha=6$ proves advantageous for sensing applications, albeit at the cost of a reduced generation rate.

\begin{figure}
	\subfigure[\label{ft_02}]{\includegraphics[scale=0.38]{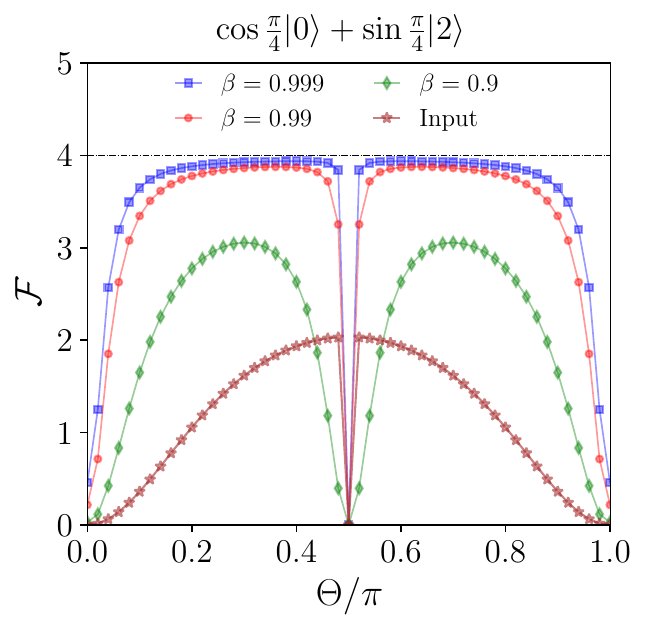}}
	\subfigure[\label{ft1_02}]{\includegraphics[scale=0.38]{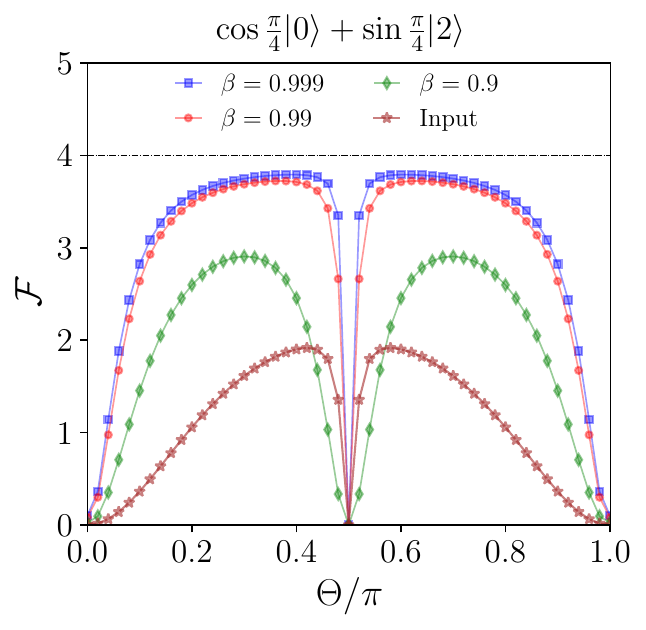}}
	\caption{Sensing the phase $\Theta$ in the operation $e^{\mi\Theta\hat{n}}$ (\fleq{eq:sem}), using filtered optical states from the single-round cavity-QED setup. $\beta$ represents  cavity efficiency used to generate the states. The line ($\mathcal{F}=4$) shows the QFI of ideal states. \subref{ft_02} Displaced squeezed vacuum state $\rho(1.035,0.247,0)$ is used as an input to filter the state $\cos (\pi/4) \ket{0}+\sin (\pi/4)\ket{2} $. The atomic success rates for the cavity efficiencies $\qty{0.999,0.99,0.9} =\qty{0.515,0.515,0.525}$.
\subref{ft1_02} Effect of noisy detection on the filtered state $\cos(\pi/4)\ket{0} + \sin(\pi/4)\ket{2}$. Projective measurements with $0.97$ efficiency are considered, while other parameters match those in Fig.~\subref{ft_02}.} \label{fg:fs02}
\end{figure}
The phase sensing capabilities are analyzed using filtered states $\ket{\theta} = \cos\theta \ket{0} + \sin\theta \ket{2}$, the sensing operation in Eq.~\eqref{eq:sem}, the Fisher information from Eq.~\eqref{eq:fi}, and projective measurements of the form $\ket{\phi} = \cos\phi\ket{0} + \sin\phi\ket{2}$.
Similarly, as for the sensing with Fock states, we project on the same class of states we generate \cite{McCormick_2019}.
As demonstrated in Fig.~\ref{ft_02}, balanced filtered states exceed the sensing capabilities of input displaced squeezed states and also any superposition up to single photon state having $\mathcal {F}=1$, despite substantial cavity losses ($\beta = 0.99$). Furthermore, they approach the quantum Fisher information bound ($\mathcal{F} = 4$) for loss values up to $\beta = 0.99$. 
Fig.~\ref{ft1_02} demonstrates detector loss reduces the Fisher information, yet the system nearly attains the QFI limit ($\mathcal{F} = 4$) despite $0.97$ loss. Collectively, the results in Fig.~\ref{fg:fs02} establish cavity QED as a promising platform for generating of Fock superposition states for phase sensing.

\section{conclusion}
The generation of quantum non-Gaussian states at optical frequencies has long been a desirable goal for applications in communication, sensing and recently, quantum error correction for quantum computing. While various protocols have been proposed, most have been limited to the heralded generation of such states using TMSV from non-linear optics and PNRDs with still non-unity efficiency.

As a viable alternative, we explored the merits of the cavity-QED filtration protocol, which enables the generation of non-Gaussian states in travelling pulses
with a success rate of $20\%$, by an order higher than conventional methods based on TMSV and PNRD, without compromising the quality of photon number statistics.  Our scheme allows using available coherent states of light to herald high-quality large Fock-states, which is challenging for conventional methods. We assessed the advantages of the filtered state $\ket{10}$ using a hierarchy of QNG criteria, bunching capability, and Fisher information for sensing of the optical displacement and squeezing.

As an attractive high target, our study shows that input squeezed coherent states, with optimized parameters and two atomic measurements, can filter the state $\ket{10}$ under realistic atom-cavity conditions. The filtered state has QNG-depth of $T=0.97$, which is remarkably close (by 75\%) to the ideal Fock-state depth of $T=0.96$. Additionally, the bunching capability analysis confirms that the filtration protocol produces noise tolerant Fock-states with minimal contributions from higher Fock-states.
Such states can be used for displacement and squeezing sensing, which requires high-quality sub-Poissonian light.

Beyond generating non-Gaussian states at each iteration, filtration also serves as a powerful tool for quantum state engineering. While the process is inherently limited in its ability to adjust coefficients, this constraint can be mitigated by optimizing input displaced squeezed vacuum states. This allows for the creation of binary superposition of Fock states up to two photons, having coherence QNG-depth of $T=0.875$, which is nearly identical (by 92.5\%) to the ideal superposition state depth of $T=0.865$.
Remarkably, such QNG superposition states can be used for optical phase sensing to approach the ultimate Heisenberg limits, as demonstrated for mechanical states \cite{McCormick_2019} can be further extended after the proof-of-principle experimental verifications.

In Appendix.~\ref{ap:tam}, we compare the filtration protocol using a single atom with that using two atoms. Introducing two atoms maintains crucial parameters $ \qty{C,\beta }$ unchanged, offering no significant advantages over single-atom filtration for the generation of Fock-states. Although a slight advantage may arise in the two-atom case due to stronger interaction ($\sqrt{2}g$), increasing $g$ in the single-atom scenario yields results similar to those in the two-atom scenario.

An alternative technique for the deterministic generation of Fock-states involves controlling atom-cavity interactions \cite{1996_law}. However, this control becomes cumbersome, and the fidelity of the generated Fock-states is time dependent and decreases with higher photon-number states \cite{2020uria,Cosacchi_2020}. Another approach to generate Fock-states utilizes atom-cavity evolutions and qubit measurements, but this method forms the Fock-states as standing waves rather than travelling pulses \cite{Wang_2017,zhang2024,Magro_2023}. This approach has recently been demonstrated in superconducting circuits using dispersive Hamiltonians ($\hat{\sigma}_z \hat{a}^\dagger \hat{a}$) \cite{deng2023}.

This work simultaneously employs different concepts of QNG rank and depth, bunching, and sensing capabilities to establish a definitive framework for optimizing optical cavity QED protocols
experiments for propagating pulses. By this, they will achieve the same conclusively justified quality as mechanical \cite{podhora2022,rahman2025,kov2024} and microwave intracavity experiments \cite{krisnanda2025}, which have already been tested for their QNG rank.
By analyzing the necessary conditions for atomic (or quantum dot) cavity QED systems \cite{Cosacchi_2020,Najer_2019,Hacker_2019,Hacker_2016} alongside available optical delay elements, we highlight the feasibility of the sequential filtration protocol in generating non-Gaussian states in traveling optical pulses. This approach not only surpasses the limitations of photon-number-resolving detectors on the Gaussian states but
also provides the first conclusive justification for a feasible optical cavity QED protocol capable of producing high-rank optical Fock states and their superpositions. These results offer a valuable tool for advancing quantum sensing and computing technologies.

\begin{acknowledgments}
We would like to thank Kimin Park for discussion about quantum sensing.
G.P.T. acknowledges the project of 21-13265X, and L.L. acknowledges the grant 23-06015O, both from the Czech Science Foundation. R.F. was supported by the European Union’s HORIZON Research and
Innovation Actions under Grant Agreement no. 101080173 (CLUSTEC) and the Quantera project CLUSSTAR (8C24003) of
MEYS, Czech Republic. Project CLUSSTAR has received funding from the European Union’s Horizon
2020 Research and Innovation Programme under Grant Agreement No.731473 and 101017733 (QuantERA).
R.F. was also funded by a grant from the Programme Johannes Amos Comenius under the Ministry of Education, Youth and Sports of the Czech Republic, number CZ.02.01.01/00/22\_008/0004649 (QUEENTEC). 
As set out in the Legal Act, beneficiaries must ensure that the open access to the published version or the final peer-reviewed manuscript accepted for
publication is provided immediately after the date of publication via a
trusted repository under the latest available version of the Creative
Commons Attribution International Public Licence (CC BY) or a licence
with equivalent rights. For long-text formats, CC BY-NC, CC BY-ND, CC
BY-NC-ND or equivalent licenses could be applied.
\end{acknowledgments}

%\bibliography{daf.bib}
%\bibliographystyle{apsrev4-2}

%apsrev4-2.bst 2019-01-14 (MD) hand-edited version of apsrev4-1.bst
%Control: key (0)
%Control: author (72) initials jnrlst
%Control: editor formatted (1) identically to author
%Control: production of article title (-1) disabled
%Control: page (0) single
%Control: year (1) truncated
%Control: production of eprint (0) enabled
%

\appendix
\onecolumngrid
\section{Thermal state filtration} 
\subsection{Success rates and delay-loop noise} \label{ap:srno}
Although squeezed coherent and thermal states yield similar Fock-states the success probability, sucessr rate (atomic measurements) varies depending on the input states states as shown in Table.~\ref{tab:mes}.
\begin{table*}[!tbh]
\begin{ruledtabular}
\begin{tabular}{c|c|c|c|c|c|c}
\diagbox{$N$}{Input} & $\rho(\sqrt{10},0,0)$ & $\rho(\sqrt{6},0,0)$ & $\rho(0,0,3)$ & $\rho(\sqrt{10},0.5,0.0485)$ & $\rho(\sqrt{4.5},0.5,0.04)$ & $\rho(\sqrt{50},0,0)$\\
\colrule
1&0.50 & 0.50 & 0.57 & 0.50 & 0.50 & 0.50 \\ 
2&0.50 & 0.50 & 0.36 & 0.50 & 0.53 & 0.50 \\ 
3&0.52 & 0.36 & 0.76 & 0.68 & 0.89 & 0.50 \\ 
4&0.90 & 0.40 & 0.09 & - & -& 0.50 \\ 
5&- & - & - & - & - & 0.75 \\ \hline
$\mathcal{P}$ &0.12 & 0.04 & 0.01 & 0.17  & 0.236 & 0.047
\end{tabular}
\end{ruledtabular}
\caption{\label{tab:mes}%
Comparison of atomic measurements in the filtration protocol for various Gaussian input states. $N$ denotes the measurement after $n^{\text{th}}$  reflection  and $\mathcal{P}$ is the product of atomic measurements. $\rho(\alpha,r,\bar{n})$ denotes the displaced squeezed thermal state in \fleq{eq:ins}. The atom-cavity  parameters are set to $\qty{C,\beta}=\qty{250,0.99}$, see Fig.~\ref{inp}.}
\end{table*}

From the atomic measurement probabilities, it is evident that the squeezed coherent states are beneficial. Though shifted coherent states exhibit lower success rates, they have the advantage of low contribution from higher Fock-states i.e. $\expval{\rho}{11} \approx 0$.
The thermal states demonstrate that the coherence (off-diagonal) in the input light state is irrelevant, and the filtration of Fock states is immune to dephasing-noise.

\begin{figure}[!tbh]
	\centering
	\includegraphics[scale=0.6]{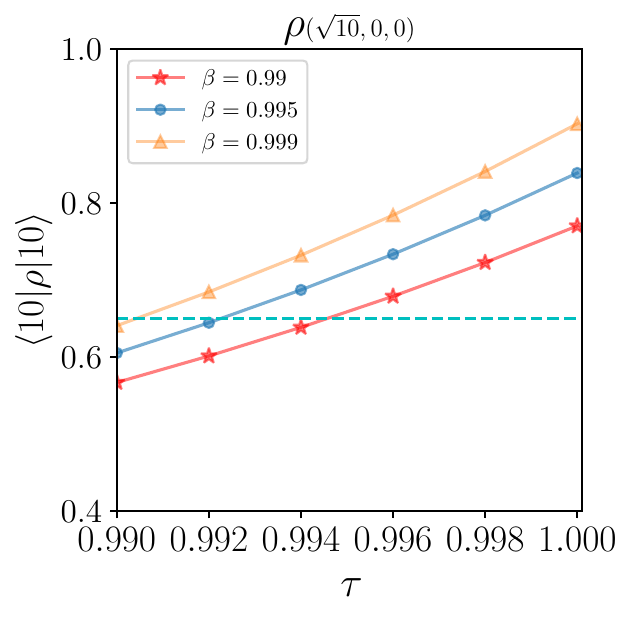}
	\caption{Delay loss effects on state filtration. The state $\ket{10}$ is filtered from the coherent state $\rho(\sqrt{10},0,0)$. Delay loss is modeled as a beam splitter with transmissivity $\tau$. It is clear that the losses ($\beta=0.999$, $\tau=0.994$), ($\beta=0.995$, $\tau=0.996$) and ($\beta=0.99$, $\tau=0.998$) are equivalent for filtration.}\label{fig:ui}
\end{figure}

Moreover, residual small losses in the loop only affect the optical states, not light-matter interaction.
In general, this will further reduce the efficiency of the protocol because, intuitively, it can be viewed as increasing the scattering losses $\kappa_l$ (see Fig.\ref{fig:ui}), which reduces both the cooperativity $C$ and the cavity efficiency $\beta$.

\section{Two mode squeezed states}\label{ap:tmsv}
Here we derive probability and fidelity of Fock-states generated using TMSV and PNRDs.
In order to have a faithful comparison and to accommodate cavity scattering losses we pass both  the modes through a beam splitter with transmittances ($\beta$) identical to cavity losses.
We  consider a TMSV state generated by the action of two mode squeezing operator on two mode vacuum state as follows:
  %The TMSV state can be written as follows in the Fock basis: 

\begin{align} \label{twomodesq}
   \begin{aligned}
   \ket{\psi}_{AB} & = \exp[r (\hat{a}_A^{\dagger} \hat{a}_B^{\dagger}-\hat{a}_A
  		\hat{a}_B) ]|0\rangle_A |0\rangle_B \\
&= 
  	  \sqrt{1-\lambda^2} \sum_{n=0}^{\infty}(-\lambda)^n 
  	|n\rangle_A |n \rangle_B,
  \end{aligned}
  \end{align}
where $A,B$ denote two-modes and 
$\lambda = \tanh r$.  To generate Fock-state $|n\rangle$, we perform $n$-photon detection on one of the modes of the   TMSV state.  
%As earlier,  we perform $n$-photon detection on one of the modes of the decayed TMSV state to generate   Fock-state $|n\rangle$. 
The probability of $n$-photon detection is given by
  \begin{equation}
  	\mathit{P} = \frac{\left(1-\lambda ^2 \right)   }{ \left(\lambda ^2 (\beta -1)+1\right)} \left(\frac{\lambda ^2 \beta }{ \lambda ^2 (\beta -1)+1}\right)^n.
  \end{equation}
For low loss i.e. $\beta \approx 1$, the probability of $n$-photon detection reduces to $(1-\lambda^2) \lambda^{2n}$ and for $n=10, P_\text{max} = 0.036 $ at $\lambda=\tanh(1.85) $.

Due to the decay of the TMSV state, the generated state is not identical to Fock-state $|n\rangle$. 
 The fidelity of the generated state with the state $|n\rangle$ is

\begin{align}
	F =\frac{1}{P} \frac{ (1- \lambda ^2 ) }{ \mathcal{D}}	 \qty(\dfrac{\lambda ^2 \beta		^2}{ \mathcal{D}} )^n
P_n \qty[\dfrac{\lambda ^2 (\beta -1)^2+1}{\mathcal{D}} ],
\end{align}
where $\mathcal{D} = 1-\lambda ^2 (\beta-1)^2$ and $P_n(x)$ is $n^{\text{th}}$-Legendre polynomial.

\section{Phase operation with two-atoms} \label{ap:tam}
\begin{figure*}[!tbh]
\includegraphics[scale=0.4]{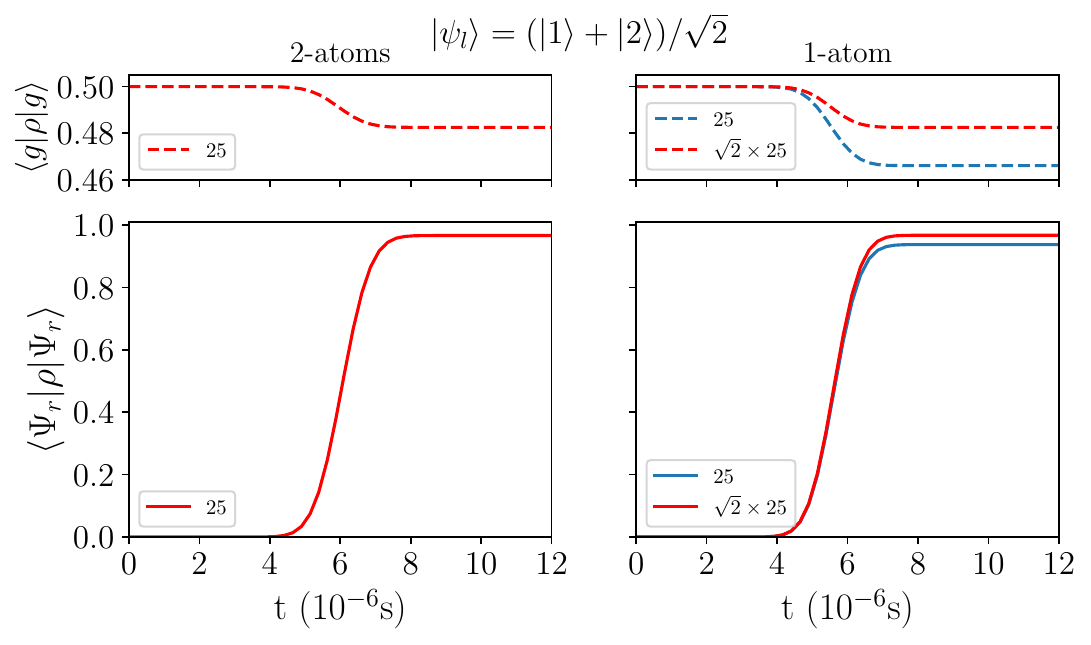}
\includegraphics[scale=0.4]{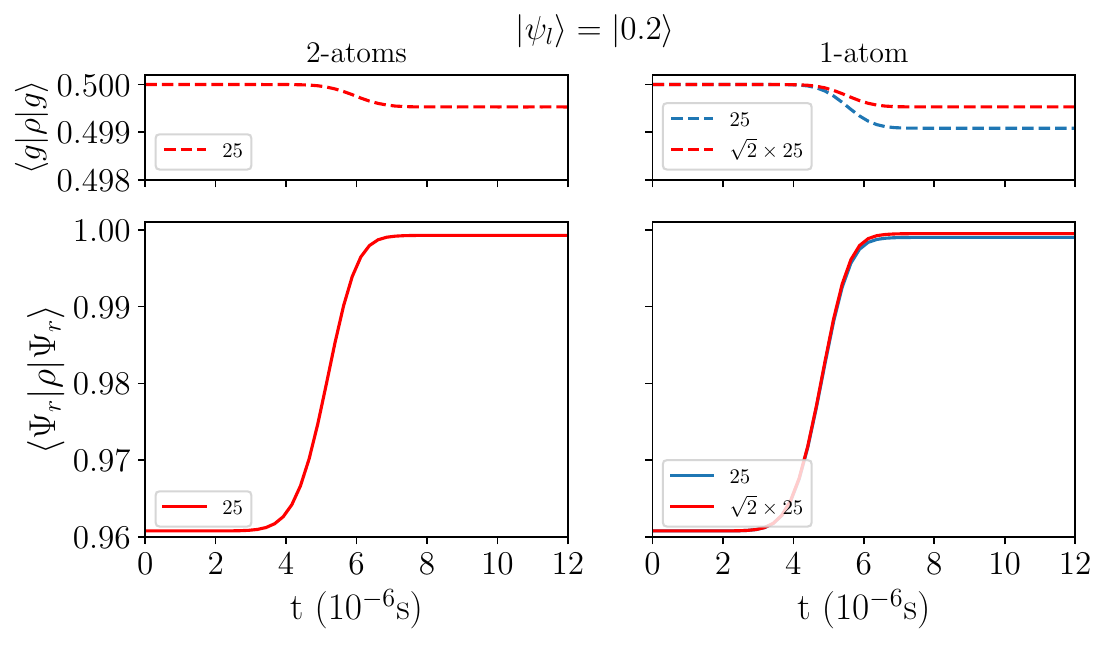}
\caption{Comparison of atom-cavity phase operation $U(\phi)$ using two atoms with one atom.$\qty{g,\kappa ,\gamma} = 2\pi\qty{25,10,3}$MHz The initial states of light are  $\ket{\psi_l}= \dfrac{1}{\sqrt{2}}(\ket{1}+\ket{2})$ and coherent state. Fidelity is calculated with state \fleq{eq:2amr}
}\label{atoms}
\end{figure*}
In the context of generating non-Gaussian states, our focus has been on filtration with single atoms. However, we now  explore potential advantages that may emerge from employing two atoms in the process. Here we focus only the atom-cavity phase operation as that will be major change due to more atoms
It is intuitive to start the atoms in the state
$
\frac{ \ket{g}+ \ket{s}}{\sqrt{2}} \otimes 
\frac{ \ket{g}+ \ket{s}}{\sqrt{2}}
$
since atoms are indistinguishable. The $\ket{gg}$, $\ket{gs}$, and $\ket{sg}$ states will all interact with the light. Only the $\ket{ss}$ state is forbidden. To keep it analogous with the single-atom scenario, we begin by initializing the atom-light system in a state.
\begin{align}
\ket{\Psi} = \dfrac{1}{2} (\ket{gg}+ \ket{ss}) \otimes \ket{\psi_l},\label{eq:2am}
\end{align}
and atom cavity reflection will result in
\begin{align}\label{eq:2amr}
\ket{\Psi_r} = \dfrac{1}{\sqrt{2}} (\ket{gg} \ket{\psi_l} + \ket{ss} e^{\mi \pi \hat{n} } \ket{\psi_l}).
\end{align}
Introducing two atoms indeed changes the atom-cavity coupling strength as $g^2 \to 2g^2$ and the total spontaneous decay rate to $2\gamma$. However, crucial factors like $C=\frac{g^2}{\kappa \gamma}$ and $\eta = \frac{\kappa_l}{\kappa}$ remain unaffected, offering no significant advantages over single atom filtration for the Fock-states.

It is noteworthy that \fleq{eq:2amr} is derived under the assumption $\pdv{\sigma_{gg}}{t} = 0$, i.e., the atomic population is unchanged. From this standpoint, potential advantages with two atoms may emerge to maintain the atomic population. 
In Fig.~\ref{atoms}, we compare filtration outcomes with one and two atoms for parameters $\qty{g,\kappa ,\gamma} = 2\pi \qty{25,10,3}$,
We perform numerical simulations utilizing virtual cavities \cite{2019_inpout}, 
to determine the necessary quantities: $\ev{\rho}{gg}$, $\ev{\rho}{g}$ and
compare the fidelities ($\ev{\rho}{\Psi_r}$) with \fleq{eq:2amr}.

The input state $\ket{\psi_l} = (\ket{0}+\ket{1})/{\sqrt{2}}$ gives fidelities of $0.97$ (2-atoms) and $0.93$ (1-atom), primarily due to slight excitation to $\ket{e}$ i.e. $\pdv{\ev{\rho}{g}}{t} \neq 0$ and $\pdv{\ev{\rho}{gg}}{t} \neq 0$. This excitation can be mitigated by increasing $g$. For instance, with $g=\sqrt{2}\times 25$, excitation to $\ket{e}$ decreases, and the fidelity is similar to that of two atoms. While the coherent state ($\ket{\psi_l}=\ket{\sqrt{0.2}}$) gives similar results,  due to its low amplitude the differences in fidelities are negligible.

\section{Phase operation with dephasing}
\begin{figure*}[!tbh]\label{fig:apde}
\subfigure[\label{dep_al}] {\includegraphics[scale=0.45]{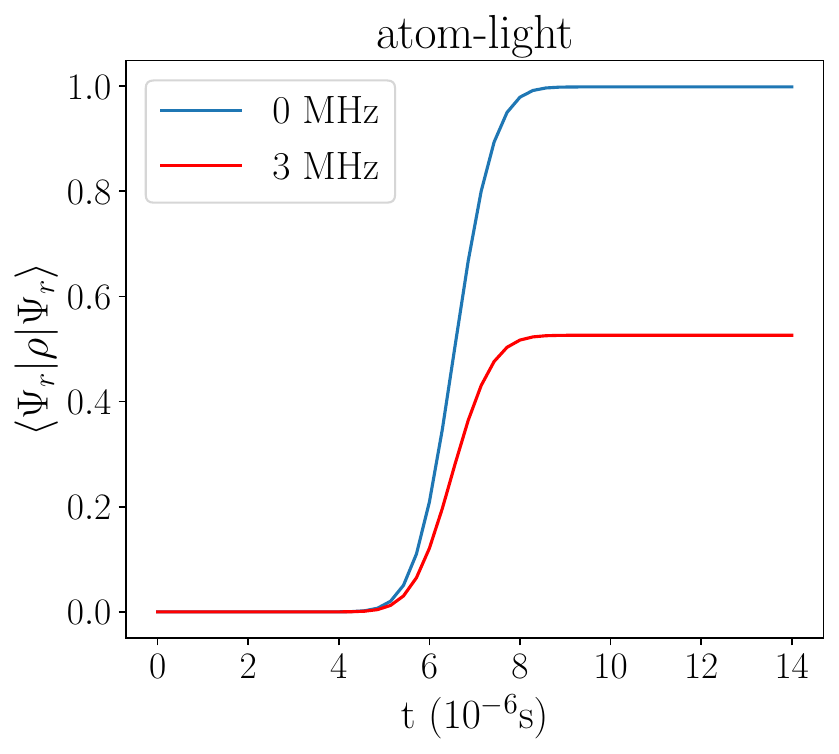}}
\subfigure[\label{de0}] {\raisebox{1cm}{\includegraphics[scale=0.55]{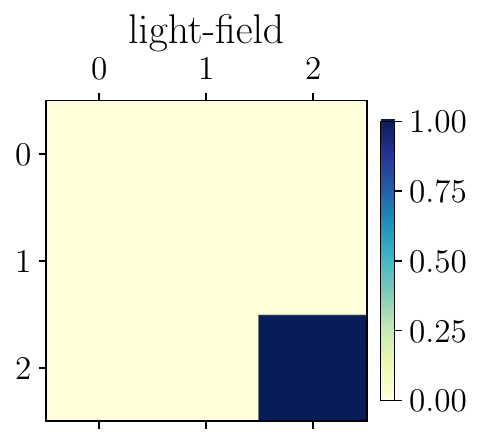}}}
\subfigure[\label{de3}] {\raisebox{1cm}{\includegraphics[scale=0.55]{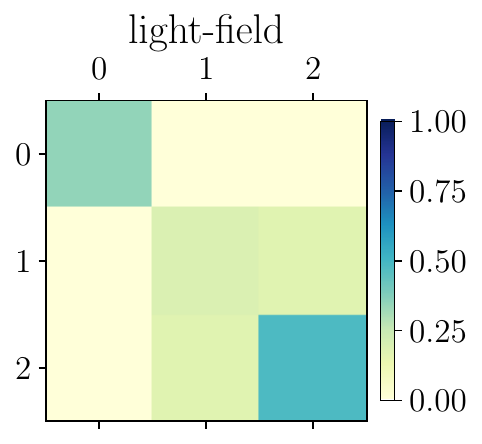}}}
\caption{Comparison of the atom-cavity phase operation $U(\phi)$ with dephasing. \subref{dep_al} shows the phase operation under dephasing noise with $\gamma_d = 2\pi \times \qty{0, 3}$ MHz. The initial state of the light field is $\frac{1}{\sqrt{2}}(\ket{1}+\ket{2})$, and fidelity is calculated with the state \fleq{ap:1am}. \subref{de0} ($\gamma_d = 0$) and \subref{de3} ($\gamma_d = 2\pi \times 3$ MHz) display the states of the light field at $\text{t} = 14 \times 10^{-6}$ s after the measurement on the atom. Other atom-cavity parameters are set to $\qty{g, \kappa, \gamma} = 2\pi \times \qty{25, 10, 0}$ MHz.
}\label{atoms}
\end{figure*}

Here, we study the effect of dephasing during the atom-cavity phase operation $U(\phi)$. Although this type of noise is common in superconducting circuits \cite{deng2023}, in typical optical resonators, it can occur due to jittering of the cavity mirrors \cite{2012_molmer}.
As above we simulate the phase operation using virtual cavities along with dephasing noise 
$
\mathcal{D} (\rho) =  \gamma_d/2 \qty[ 2 \hat{n} \rho \hat{n}^\dagger -  \hat{n}^2 \rho - \rho \hat{n}^2 ],
$
added during dynamical evolution of atom-cavity system. For the single atom \fleq{eq:2amr} modifies to
\begin{align}\label{ap:1am}
\ket{\Psi_r} = \dfrac{1}{2} \qty[ (\ket{g} - \ket{s}) \otimes \ket{1} + (\ket{g}+\ket{s}) \otimes \ket{2}  ].
\end{align}
Here, the initial state of the light field is given by $\ket{\psi_l} = \frac{1}{\sqrt{2}}(\ket{1} + \ket{2})$. In the absence of dephasing noise, measuring the atom in the $\ket{g} + \ket{s}$ basis should project the light field onto the state $\ket{2}$. In Fig.\ref{dep_al}, we compare the phase operation with and without dephasing. As expected, the unitary operation without dephasing outperforms the noisy operation. Furthermore, with atomic measurement, the ideal scenario ($\gamma_d = 0$ MHz ) results in the state $\ket{2}$, as shown in Fig.\ref{de0}. However, when dephasing noise is introduced, it disrupts the phase operation, resulting in a mixed state after filtration, as illustrated in Fig.~\ref{de3}.

\section{Squeezed coherent state} \label{ap:scs}
The displaced squeezed vacuum state can be written as $\ket{\alpha e^{\mi \phi},r e^{\mi \theta}}= \hat{\mathcal{D}} (\alpha e^{\mi \phi} ) \hat{\mathcal{S}} (r e^{\mi \theta}) \ket{0}$
\begin{align}\label{aeq:var}
\begin{aligned}
\ket{\alpha e^{\mi \phi},r e^{\mi \theta}}= & \dfrac{e^{-\frac{\alpha^2}{2} (1+ e^{\mi (\theta-2\phi)} \tanh r)}}{\sqrt{\cosh r}} 
\sum_{n} 
\dfrac{(\frac{1}{2} e^{\mi \theta} \tanh r)^{n/2}}{\sqrt{n!} }  
 H_n\qty[\dfrac{\alpha (e^{\mi \phi} \cosh r + e^{\mi (\theta-\phi) } \sinh r)}{\sqrt{e^{\mi \theta} \sinh{2r}}}] \ket{n},\\
\ket{\alpha e^{\mi \phi},r e^{\mi 2\phi}} = & \dfrac{e^{-\frac{\alpha^2}{2} (1+ \tanh r)}}{\sqrt{\cosh r}} 
\sum_{n} \sqrt { \dfrac{(\tanh r)^{n}}{  2^n n! } } 
e^{\mi n \phi}  H_n\qty[\dfrac{\alpha e^{r} }{\sqrt{ \sinh{2r}}}]\ket{n},\\
p_n \equiv & \abs{\braket{n}{\alpha e^{\mi \phi},r e^{\mi 2\phi}}}^2  =  p_0 \dfrac{(\tanh r)^n}{n! ~2^n}  \abs{H_n\qty[\dfrac{\alpha e^{r}}{\sqrt{\sinh{2r}}}]}^2,
\end{aligned}
\end{align}
where $p_0 = \frac{e^{-\alpha^2 (1+\tanh r)}}{\cosh r} $ and $H_n$ represents the $n^{\text{th}}$ order Hermite polynomial .  $p_n \equiv \abs{\braket{n}{\alpha e^{\mi \phi},r e^{\mi 2\phi}}}^2 $ are $n^\text{th}$-photon probabilities.
On expanding \fleq{aeq:var} gives:

\begin{align}
\ket{\alpha e^{\mi \phi},r e^{\mi 2 \phi}} = C_0 \qty[ \ket{0}  + 
e^{\mi \phi} C_1\ket{1}+
e^{\mi 2\phi}  C_2\ket{2}+  e^{\mi 3\phi}  C_3\ket{3}+
\sum_{n=4} \mathcal{O}_n \ket{n}].
\end{align}

To improve the fidelity of the output states we choose {$\alpha,r$} such that $\sum_{n=0}^3 p_n \approx 1$ thus we can drop the higher order terms and rewrite 
\begin{align}
\psi_\text{sc} \approx C_0 \qty[ \ket{0}  + C_1\ket{1}+C_2\ket{2}+  C_3\ket{3} ].
\end{align}
after the interaction with an atom in the cavity the atom-light state is written as
\begin{align}
\psi_\text{sc} \approx C_0 \qty[ \ket{0}  + C_1\ket{1} + C_2\ket{2} +  C_3\ket{3} ] \ket{g} ~+~
C_0 \qty[ \ket{0}  - C_1\ket{1} + C_2\ket{2} -  C_3\ket{3} ]  \ket{s},
\end{align}
and Hadamard operation gives
\begin{align}
\psi_\text{sc} \approx C_0 \qty[ \ket{0}  + C_2\ket{2}  ] \ket{g}+
C_0 \qty[  C_1\ket{1} + C_3\ket{3} ] \ket{s},
\end{align}
and measuring the atoms in the state $\ket{g}$ gives
$
\psi_{02} \propto C_0 \qty[ \ket{0}  + \tan \theta \ket{2}  ] ,
$
where $\theta = \tan^{-1}\qty({ \dfrac{\tanh r}{2\sqrt{2}}  H_2\qty[\dfrac{\alpha e^{r} }{\sqrt{ \sinh{2r}}}]})$.
All numerical calculations are performed with states in \fleq{aeq:var} using sufficiently large basis.

\section{Fisher information for phase sensing} \label{ap:fs02}
\begin{table*}
	\setlength{\extrarowheight}{4pt} % Add extra space around each row
	\renewcommand{\arraystretch}{1.5} % Increase row height by 50%
	\begin{ruledtabular}
		\begin{tabular}{c|c|c}
			state &  FI ($\mathcal{F}$) & $\expval{\hat{n}}$ \\
			\colrule 
			$\cos \theta \ket{0} + \sin\theta \ket{N}$ & QFI = $N^2 \sin^2 2\theta $ & $N \sin^2 \theta$  \\ 
$\cos \theta \ket{0} + \sin\theta \ket{1}$ & CFI=$\frac{ \sin^2(2\theta)  \sin^2(\Theta)}{1-\sin^2(2\theta)\cos^2(\Theta)}$ & $  \sin^2 \theta $  \\
			$ \cos \theta \ket{0} +  \sin\theta \ket{2}$ & CFI=$\frac{4 \sin^2(2\theta)  \cos^2(\Theta)}{1-\sin^2(2\theta)\sin^2(\Theta)}$ & $ 2 \sin^2 \theta $  		  
		\end{tabular}
	\end{ruledtabular}
	\caption{\label{tab:fish}%
		Comparison of Fisher information and average photon numbers for states $\ket{\theta}=\cos \theta \ket{0} +  \sin\theta \ket{N}$.}\label{ap:tb3}
\end{table*}
The Quantum fisher information for pure states under the action $e^{\mi \Theta \hat{n} }$  is simplified as 
\begin{align}
	\mathcal{F} =  4 [\expval{\hat{n}^2}- \expval{\hat{n}}^2].
\end{align}
It is clear from Table~\ref{ap:tb3} that the QFI can be achieved using binary projections with $\{\dyad{\pi/4}, I - \dyad{\pi/4}\}$ and balanced superposition states ($\theta = \pi/4$). Furthermore, superpositions up to $\ket{1}$ have a maximum CFI of one, whereas states with $\ket{0}$ and $\ket{2}$ superpositions yield four times the CFI.
For a general state with POVM measurements $\Pi_m$, the Fisher information is expressed as
\begin{align}
\mathcal{F}_\Theta = \sum_{m} \dfrac{1}{p(m|\Theta)} \qty(\pdv{\Theta}  p(m|\Theta))^2 \quad \text{with} \quad
p(m|\Theta) = \text{Tr}[\Pi_m e^{-\mi \Theta \hat{n}} \rho e^{\mi \Theta \hat{n}}],
\end{align}
where $\rho$ is the filtered state from the cavity. Using the state $\ket{\theta} = \cos \theta \ket{0} + \sin \theta \ket{2}$ and the projective measurements $\Pi_1 = \dyad{\phi}$ and $\Pi_2 = \mathrm{I} - \Pi_1$, the probabilities are given as
\begin{subequations}
\begin{align}
p_1(\Theta) \equiv \abs{\braket{\theta}{\phi}}^2 &= \cos^2 \theta \cos^2 \phi + \sin^2 \theta \sin^2 \phi + \frac{1}{2} \sin(2\theta) \sin(2\phi) \cos(2\Theta),\\
	F(\Theta) &= \dfrac{\sin^2(2\theta) \sin^2(2\phi) \sin^2(2\Theta)}{p_1(\Theta,\theta,\phi)p_2(\Theta,\theta,\phi)},
\end{align}
\end{subequations}
where we used $p_2 (\Theta)= 1-p_1(\Theta,\theta,\phi)$. Note that $p_1(\Theta,\theta,\phi)$ is symmetric with respect to $\theta$ and $\phi$. For a given $\Theta$, the optimized Fisher information can be obtained with  $\phi=\theta=\pi/4$. 
%However, the maximum Fisher information is achieved when $\phi=\theta=\pi/4$.

\end{document}